\documentclass[letterpaper,twocolumn,10pt]{article}
\usepackage{usenix,endnotes,epsfig}

\usepackage{amssymb}
\usepackage{amsmath}
\usepackage{amsfonts}
\usepackage{url,color,fancyvrb}
\usepackage{times}
\usepackage{subfigure}
\usepackage{array}
\usepackage{xspace}
\usepackage{multirow}
\usepackage{multicol}
\usepackage{program}
\usepackage{algorithm}
\usepackage{enumitem}
\usepackage{listings}
\newcommand{\ppis}{{P3}\xspace}

\newcommand{\authnote}[2]{\ifnum\authnotes=1\begin{quote}\textbf{#1 says:} #2\end{quote}\fi}

\newcommand{\para}[1]{\vspace*{1ex}\noindent\textbf{#1}}

\newcommand{\camera}[1]{#1}

\newif\iftechrep
\techreptrue

\definecolor{brown}{cmyk}{0,0.81,1,0.60}
\definecolor{magenta}{rgb}{0.4,0.7,0}
\definecolor{gray}{rgb}{0.5,0.5,0.5}
\definecolor{red}{rgb}{1,0,0}
\definecolor{green}{rgb}{0.5,0,0.5}
\definecolor{blue}{rgb}{0,0,1}

\newcommand{\etal}{\emph{et al.}\xspace}

\newcommand{\mypar}[1]{\vspace*{0.5ex}\noindent\textbf{#1}}

\begin{document}


\title{\ppis: Toward Privacy-Preserving Photo Sharing}

\author{
{\rm Moo-Ryong Ra}\vspace{1ex}\\
\and
{\rm Ramesh Govindan}\vspace{1ex}\\
\vspace{-3ex}\\
University of Southern California
\and
{\rm Antonio Ortega}\vspace{1ex}\\
} 


\maketitle

\subsection*{Abstract}

  \camera{With increasing use of mobile devices, photo sharing
  services are experiencing greater popularity.}
%
%
  Aside from providing storage, photo sharing services enable
  bandwidth-efficient downloads to mobile devices by performing
  server-side image transformations (resizing, cropping).
  On the flip side, photo sharing services have raised privacy
  concerns such as leakage of photos to unauthorized viewers and the
  use of \camera{algorithmic} recognition technologies by providers.
  To address these concerns, we propose a privacy-preserving photo
  encoding algorithm that extracts and encrypts a small, but
  significant, component of the photo, while preserving the remainder
  in a public, standards-compatible, part.
  These two components can be separately stored.
  This technique significantly reduces \iftechrep the signal-to-noise
  ratio and \fi the accuracy of automated detection and recognition on
  the public part, while preserving the ability of the provider to
  perform server-side transformations to conserve download bandwidth
  usage.
  Our prototype privacy-preserving photo sharing system, \ppis, works
  with Facebook, and can be extended to other services as well.
  \ppis requires no changes to existing services or mobile application
  software, and adds minimal photo storage overhead.



\section{Introduction}
\label{sec:intro}

With the advent of mobile devices with high-resolution on-board
cameras, 
photo sharing has become \camera{highly} popular.
Users can share photos either through photo sharing services like
Flickr or Picasa, or popular social networking services like Facebook
or Google+.
These \emph{photo sharing service providers} (PSPs) now have a large
user base, to the point where PSP photo storage subsystems have
motivated interesting systems research~\cite{Haystack}.

However, this development has generated privacy concerns
(Section~\ref{sec:motiv}).
Private photos have been leaked from a prominent photo sharing
site~\cite{PhotobucketFusking}.
Furthermore, widespread concerns have been raised about the
application of face recognition technologies in
Facebook~\cite{FacebookFace}.
Despite these privacy threats, it is not clear that the usage of photo
sharing services will diminish in the near future.
This is because photo sharing services provide several useful
functions that, together, make for a seamless photo browsing
experience.
In addition to providing photo storage, PSPs also perform several
server-side image transformations (like cropping, resizing and color
space conversions) designed to improve user perceived latency of photo
downloads and, incidentally, bandwidth usage (an important
consideration when browsing photos on a mobile device).

In this paper, we explore the design of a privacy-preserving photo
sharing algorithm (and an associated system) that \emph{ensures photo
  privacy without sacrificing the latency, storage, and bandwidth
  benefits provided by PSPs}.
This paper makes two novel contributions that, to our knowledge, have
not been reported in the literature (Section~\ref{sec:related}).
%
\camera{First, the}
design of the \ppis algorithm (Section~\ref{sec:approach}),
  which prevents leaked photos from leaking \emph{information}, and
  reduces the efficacy of automated processing (e.g., face detection,
  feature extraction) on photos, while still permitting a PSP to apply
  image transformations.  It does this by splitting a photo into a
  public part, which contains most of the \emph{volume} (in bytes) of
  the original, and a secret part which contains most of the
  original's \emph{information}.
%
\camera{Second, the} 
design of the \ppis system (Section~\ref{sec:sysarch}),
  which requires no modification to the PSP infrastructure or
  software, and no modification to existing browsers or
  applications. \ppis uses interposition to transparently
  encrypt images when they are uploaded from clients, and
  transparently decrypt and reconstruct images on the recipient side.

Evaluations (Section~\ref{sec:eval}) on \camera{four} commonly used image data
sets, as well as micro-benchmarks on an implementation of \ppis,
reveal several interesting results.
Across these data sets, there exists a ``sweet spot'' in the
parameter space that provides good privacy while at the same time
preserving the storage, latency, and bandwidth benefits offered by
PSPs.
At this sweet spot, 
\iftechrep
the public part of the image has low PSNR and
\fi
algorithms like edge detection, face detection, \camera{face recognition}, 
and SIFT feature extraction are completely ineffective; 
\emph{no} faces can be detected \camera{and correctly recognized}
from the public part, \emph{no} correct features can be extracted, and
a very small fraction of pixels defining edges are correctly
estimated.
\ppis image encryption and decryption are fast, and it is able to
reconstruct images accurately even when the PSP's image
transformations are not publicly known.

\ppis is proof-of-concept of, \camera{and a step towards}, easily
deployable privacy preserving photo storage.
Adoption of this technology will be dictated by economic incentives:
for example, PSPs can offer privacy preserving photo storage as a
premium service offered to privacy-conscious customers.

\section{Background and Motivation}
\label{sec:motiv}

The focus of this paper is on PSPs like Facebook, Picasa, Flickr, and
Imgur, who offer either direct \emph{photo sharing} (e.g., Flickr,
Picasa) between users or have integrated photo sharing into a social
network platform (e.g., Facebook).
In this section, we describe some background before motivating
privacy-preserving photo sharing.

\subsection{Image Standards, Compression and Scalability}
Over the last two decades, several standard image formats have been
developed that enable interoperability between producers and consumers
of images.
Perhaps not surprisingly, most of the existing PSPs like Facebook,
Flickr, Picasa Web, and many websites~\cite{JPEG-Usage1,
  JPEG-Usage2, JPEG-Usage3} primarily use the most prevalent 
of these standards, the JPEG (Joint Photographic Experts Group) standard.
In this paper, we focus on methods to preserve the privacy of JPEG
images; supporting other standards such as GIF and PNG (usually used
to represent computer-generated images like logos etc.) are left to
future work.

Beyond standardizing an image file format, JPEG performs lossy 
compression of images.
A JPEG encoder consists of the following sequence of steps:

\mypar{Color Space Conversion and Downsampling.} In this step, the raw
  RGB or color filter array (CFA) RGB image captured by 
  a digital camera is mapped to a YUV color space. 
  Typically, the two chrominance channels (U and V) are
  represented at lower resolution than the luminance (brightness) channel (Y)
  reducing the amount of pixel data to be encoded without significant
  impact on perceptual quality.

\mypar{DCT Transformation.} In the next step, the image is divided into
  an array of blocks, each with $8\times 8$ pixels, and the 
  Discrete Cosine Transform (DCT) is applied to each block, 
  \camera{resulting in several \emph{DCT coefficients}. The mean value
    of the pixels is called the DC coefficient. The
    remaining are called AC coefficients.} 

\mypar{Quantization.} In this step, these coefficients are quantized; this
  is the only step in the processing chain where information is lost.
  For typical natural images, information tends to be concentrated in
  the lower frequency coefficients (which on average have larger
  magnitude than higher frequency ones). For this reason, JPEG applies
  different quantization steps to different frequencies.  The degree
  of quantization is user-controlled and can be varied in order to
  achieve the desired trade-off between quality of the reconstructed
  image and compression rate. We note that in practice, images shared
  through PSPs tend to be uploaded with high quality (and high rate)
  settings.

\mypar{Entropy Coding.} In the final step, redundancy in the quantized
  coefficients is removed using variable length encoding of non-zero
  quantized coefficients and of runs of zeros in between non-zero
  coefficients.

\vspace{1ex}



%
%

Beyond storing JPEG images, PSPs perform several kinds of
transformations on images for various reasons.
First, when a photo is uploaded, PSPs statically resize the image to
several fixed resolutions.
For example, Facebook transforms an uploaded photo into a thumbnail, a
``small'' image (\camera{130x130}) and a ``big'' image (720x720).
These transformations have multiple uses: they can reduce
storage\footnote{We do not know if Facebook preserves the original
  image, but high-end mobile devices can generate photos with
  4000x4000 resolution and resizing these images to a few small fixed
  resolutions can save space\camera{.}}, improve photo access latency for the
common case when users download either the big or the small image, and
also reduce bandwidth usage (an important consideration for mobile
clients).
In addition, PSPs perform \emph{dynamic} (i.e., when the image is
accessed) server-side transformations; they may resize the image to
fit screen resolution, and may also \emph{crop} the image to match the
view selected by the user.
(We have verified, by analyzing the Facebook protocol, that it
supports both of these dynamic operations).
These dynamic server-side transformations enable low latency access to
photos and reduce bandwidth usage.
Finally, in order to reduce user-perceived latency further, Facebook
also employs a special mode in the JPEG standard, called
\emph{progressive} mode.
For photos stored in this mode, the server delivers the coefficients
in increasing order (hence ``progressive'') so that the clients can
start rendering the photo on the screen as soon as the first few
coefficients are received, without having to receive all coefficients.

In general, these transformations \emph{scale} images in one fashion
or another, and are collectively called image scalability
transformations.
Image scalability is crucial for PSPs, since it helps them optimize
several aspects of their operation: it reduces photo storage, which
can be a significant issue for a popular social network
platform~\cite{Haystack}; it can reduce user-perceived latency, and
reduce bandwidth usage, hence improving user satisfaction.



\subsection{Threat Model, Goals and Assumptions}

In this paper, we focus on two specific threats to privacy that result
from uploading user images to PSPs.
The first threat is unauthorized access to photos.
A concrete instance of this threat is the practice of \emph{fusking},
which attempts to reverse-engineer PSP photo URLs in order to access
stored photos, bypassing PSP access controls.
Fusking has been applied to at least one PSP (Photobucket), resulting
in significant privacy leakage~\cite{PhotobucketFusking}.
%
The second threat is posed by \camera{automatic recognition} technologies,
by which PSPs may be able to infer social \camera{contexts} not explicitly
specified by users.
Facebook's deployment of face recognition technology has raised
significant privacy concerns in many countries (e.g.,~\cite{FacebookFace}).


The goal of this paper is \emph{to design and implement a system that
  enables users to ensure the privacy of their photos (with respect to
  the two threats listed above), while still benefiting from the image
  scalability optimizations provided by the PSP.}

Implicit in this statement are several constraints, which make the
problem significantly challenging.
The resulting system must not require any software changes at the PSP,
since this is a significant barrier to deployment; an important
implication of this constraint is that the image stored on the PSP
must be JPEG-compliant.
For a similar reason, the resulting system must also be transparent to
the client.
Finally, our solution must not significantly increase storage
requirements at the PSP since, for large PSPs, photo storage is a
concern.

We make the following assumptions about trust in the various
components of the system.
We assume that all local software/hardware components on clients
(mobile devices, laptops etc.) are completely trustworthy, including
the operating system, applications and sensors.
We assume that PSPs are completely untrusted and may either by
commission or omission, breach privacy in the two ways described
above.
Furthermore, we assume eavesdroppers may attempt to snoop on the
communication between PSP and a client.

\section{\ppis: The Algorithm}
\label{sec:approach}

In this section, we describe the \ppis algorithm for ensuring privacy
of photos uploaded to PSPs.
In the next section, we describe the design and implementation of a
complete system for privacy-preserving photo sharing.


\begin{figure}[t]
    \centering
    \includegraphics[viewport=0 100 770 550, scale=0.30, clip=true]{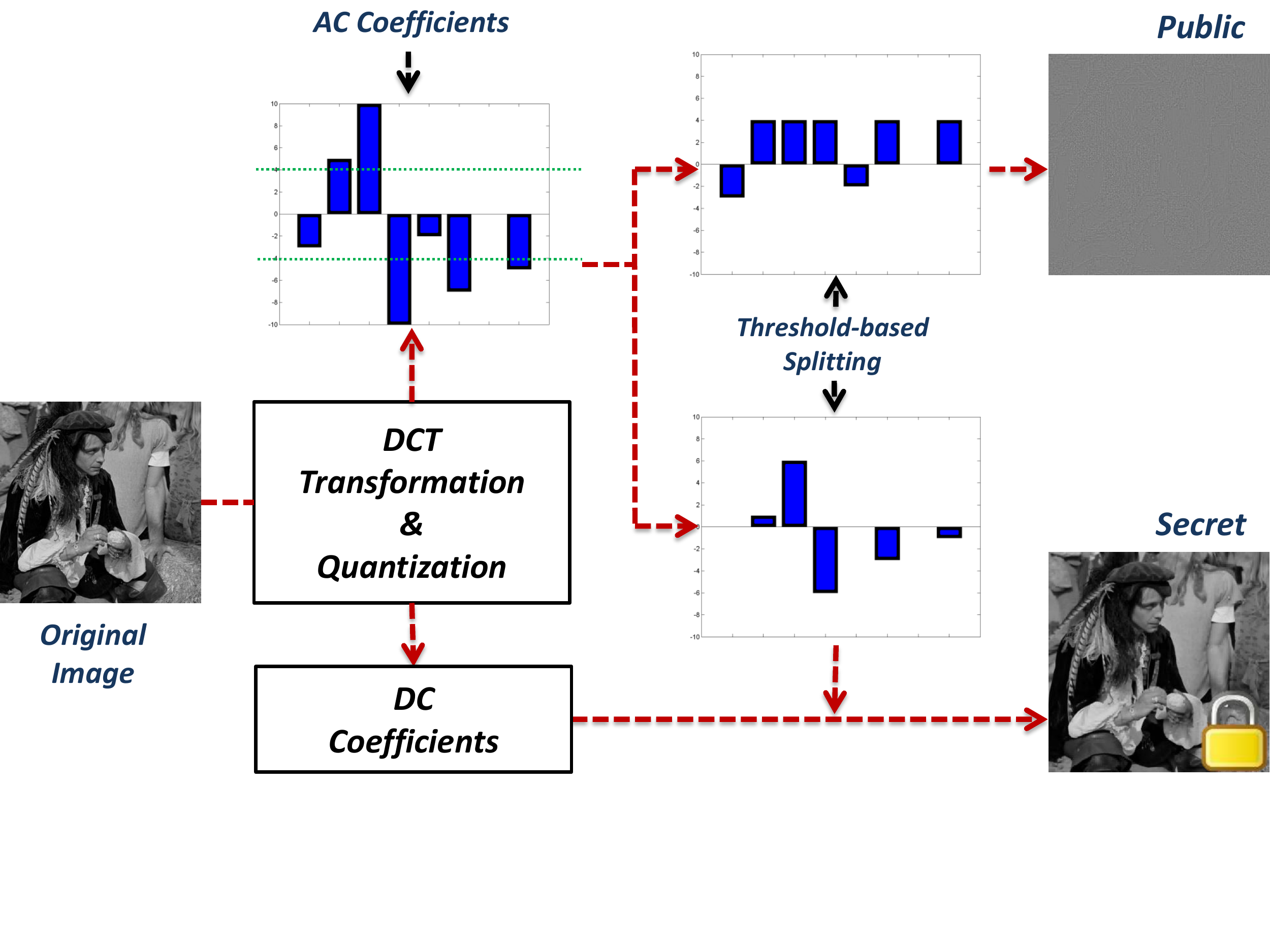}
    \caption{Privacy-Preserving Image Encoding Algorithm}
    \label{fig:algodesc}
    \vspace{-2ex}
\end{figure}

\subsection{Overview}
%
%
One possibility for preserving the privacy of photos is end-to-end
encryption.
\emph{Senders}\footnote{We use ``sender''
  to denote the user of a PSP who uploads images to the PSP.} may
encrypt photos before uploading, and \emph{recipients} use a shared
secret key to decrypt photos on their devices.
This approach cannot provide image scalability, since the photo
representation is non-JPEG compliant and opaque to the PSP so it
cannot perform transformations like resizing and cropping.
Indeed, PSPs like Facebook reject attempts to upload fully-encrypted
images.

A second approach is to leverage the JPEG image compression pipeline.
Current image compression standards use a well-known \emph{DCT
  dictionary} when computing the DCT coefficients.
A \emph{private} dictionary~\cite{K-SVD}, known only to the sender and
the authorized recipients, can be used to preserve privacy.
%
%
Using the coefficients of this dictionary, it may be possible for PSPs
to perform image scaling transformations.
However, as currently defined, these coefficients result in a non-JPEG
compliant bit-stream, so PSP-side code changes would be required in
order to make this approach work.

A third strawman approach might selectively hide faces by performing
face detection on an image before uploading.
This would leave a JPEG-compliant image in the clear, with the hidden
faces stored in a separate encrypted part.
At the recipient, the image can be reconstructed by combining the two
parts.
However, this approach does not address our privacy goals completely:
if an image is leaked from the PSP, attackers can still obtain
significant information from the non-obscured parts (e.g., torsos,
other objects in the background etc.).

Our approach on privacy-preserving photo sharing uses a
\emph{selective encryption} like this, but has a different design.
%
%
In this approach, called \ppis, a photo is divided into two
parts, a \emph{public} part and a \emph{secret} part.
The public part is exposed to the PSP, while the secret part is
encrypted and shared between the sender and the recipients (in a
manner discussed later).
Given the constraints discussed in Section~\ref{sec:motiv}, the public
and secret parts must satisfy the following requirements:
\begin{itemize}[topsep=-0.5ex,itemsep=-0.8ex, leftmargin=0.1cm]
\item It must be possible to represent the public part as a
  JPEG-compliant image. This will allow PSPs to perform image 
  scaling. 
\item However, intuitively, most of the ``important'' \emph{information} in the
  photo must be in the secret part. This would prevent attackers from
  making sense of the public part of the photos even if they were able
  to access these photos. It would also prevent PSPs from successfully
  applying recognition algorithms.
\item Most of the \emph{volume} (in bytes) of the image must reside in
  the public part. This would permit PSP server-side image scaling to
  have the bandwidth and latency benefits discussed above.
\item The combined size of the public and secret parts of the image
  must not significantly exceed the size of the original image, as
  discussed above.
\end{itemize}
%

Our \ppis algorithm, which satisfies these requirements, has two
components: a sender side encryption algorithm, and a recipient-side
decryption algorithm.

\subsection{Sender-Side Encryption}


JPEG compression relies on the \emph{sparsity} in the DCT domain of
typical natural images: a few (large magnitude) coefficients provide
most of the information needed to reconstruct the pixels.
Moreover, as the quality of cameras on mobile devices increases,
images uploaded to PSPs are typically encoded at high quality.
\ppis leverages both the sparsity and the high quality of these images.
First, because of sparsity, most information is contained in a few
coefficients, so it is sufficient to degrade a few such coefficients,
in order to achieve significant reductions in quality of the public
image.
Second, because the quality is high, quantization of each coefficient
is very fine and the least significant bits of each coefficient
represent very small incremental gains in reconstruction quality.
\ppis's encryption algorithm encode the most significant bits of (the
few) significant coefficients in the secret part, leaving everything
else (less important coefficients, and least significant bits of more
important coefficients) in the public part.
We concretize this intuition in the following design for \ppis sender
side encryption.

%

%

The selective encryption algorithm is, conceptually, inserted into the
JPEG compression pipeline after the quantization step.
At this point, the image has been converted into frequency-domain
quantized DCT coefficients.
While there are many possible approaches to extracting the most
significant information, \ppis uses a relatively simple approach.
First, it extracts the DC coefficients from the image into the secret
part, replacing them with zero values in the public part.
The DC coefficients represent the average value of each 8x8 pixel
block of the image; these coefficients usually contain enough
information to represent thumbnail versions of the original image with
enough visual clarity.

Second, \ppis uses a threshold-based splitting algorithm in which each
AC coefficient $y(i)$ whose value is above a threshold $T$ is
processed as follows:

\begin{itemize}[topsep=-0.5ex,itemsep=-0.8ex, leftmargin=0.1cm]
\item If $\lvert y(i) \rvert \leq T$, then the coefficient is represented in the public
  part as is, and in the secret part with a zero.
\item If $\lvert y(i) \rvert > T$, the  coefficient is replaced in the public
  part with $T$, and the secret part contains the magnitude of the
  difference as well as the sign.
\end{itemize}

Intuitively, this approach clips off the significant coefficients at
$T$.
$T$ is a tunable parameter that represents the trade-off between
storage/bandwidth overhead and privacy; a smaller $T$ extracts more
signal content into the secret part, but can potentially incur greater
storage overhead.
We explore this trade-off empirically in Section~\ref{sec:eval}.
Notice that both the public and secret parts are JPEG-compliant
images, and, after they have been generated, can be subjected to
entropy coding.

Once the public and secret parts are prepared, the secret part is
encrypted and, conceptually, both parts can be uploaded to the PSP (in
practice, our system is designed differently, for reasons discussed in
Section~\ref{sec:sysarch}).
We also defer a discussion of the encryption scheme 
to Section~\ref{sec:sysarch}.

\begin{figure}
    \centering
    \includegraphics[viewport=5 390 770 550, scale=0.32, clip=true]{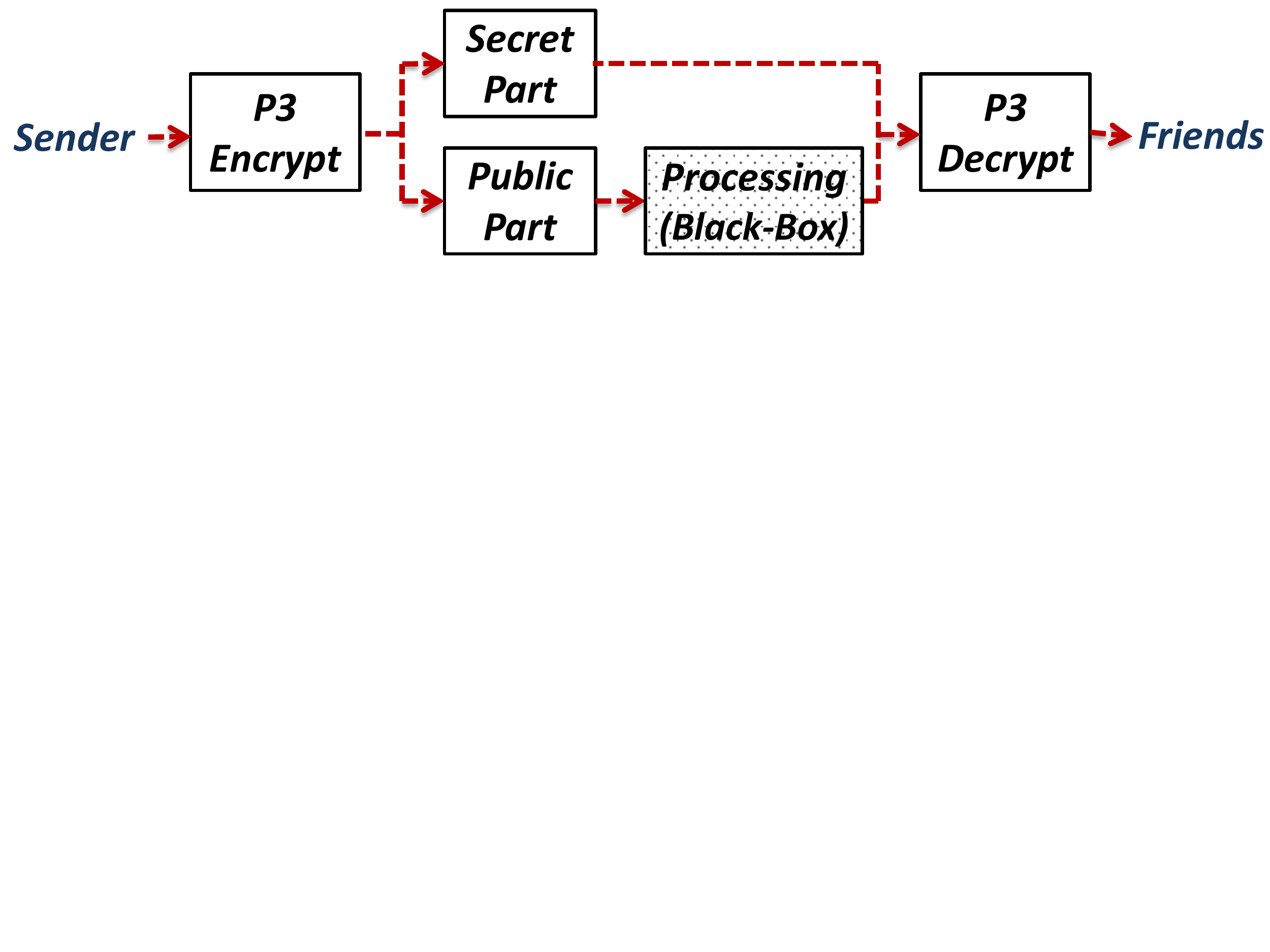}
    \vspace{-4ex}
    \caption{\ppis Overall Processing Chain}
    \label{fig:processing_desc}
    \vspace{-2ex}
\end{figure}

\subsection{Recipient-side Decryption and Reconstruction}

While the sender-side encryption algorithm is conceptually simple, the
operations on the recipient-side are somewhat trickier.
At the recipient, \ppis must decrypt the secret part and reconstruct
the original image by combining the public and secret parts.
\ppis's selective encryption is \emph{reversible}, in the sense that,
the public and secret parts can be recombined to reconstruct the original image.
This is straightforward when the public image is stored unchanged, but
requires a more detailed analysis in the case when the PSP performs some
processing on the public image (e.g., resizing, cropping, etc) in
order to reduce storage, latency or bandwidth usage. 

%

In order to derive how to reconstruct an image when the public image
has been processed, we start by expressing the reconstruction 
for the unprocessed case as a series of linear operations.

\camera{
}
Let the threshold for our splitting algorithm be denoted $T$.  
Let ${\bf y}$ be a block of DCT coefficients corresponding to a $8
\times 8$ pixel block in the
original image. Denote 
${\bf x_p}$ and ${\bf x_s}$ the corresponding DCT coefficient values 
assigned to the public and secret images,
respectively, 
for \camera{the} same block\footnote{\camera{For ease of exposition, we represent these
blocks as 64x1 vectors}}.
For example, if one of those coefficients is such that $abs(y(i)) > T$, we will have
that $x_{p}(i)= T$ and $x_s(i) = sign(y(i)) (abs(y(i)) - T)$. Since
in our algorithm the sign information 
is encoded either in the public or in the secret
part, depending on the coefficient magnitude, it is useful to
explicitly consider sign information here. 
To do so we write ${\bf x_p} = {\bf S_p} \cdot {\bf a_p}$, and ${\bf
  x_s} = {\bf S_s} \cdot {\bf a_s }$, where
\camera{
${\bf a_p}$ and ${\bf a_s }$ are absolute values of ${\bf x_p}$ and ${\bf x_s}$,
}
${\bf S_p}$ and ${\bf S_s}$ are diagonal matrices with sign information, i.e.,
${\bf S_p} = diag(sign({\bf x_p})), {\bf S_s} = diag(sign({\bf x_s}))$.
Now let ${\bf w}[i] = T$ if ${\bf S_s}[i] \neq 0$, 
where $i$ is a \camera{coefficient} index, so
${\bf w}$ marks the positions of the above-threshold coefficients.

The key observation is that ${\bf x_p}$ and ${\bf x_s}$ \emph{cannot
  be directly added} to recover ${\bf y}$ because the sign of a
coefficient above threshold is encoded correctly {\em only} in the
secret image.
Thus, even though the public image conveys sign information for that
coefficient, it might not be correct.
As an example, let $y(i) < -T$, then we will have that $x_{p}(i)= T$
and $x_s(i) = - (abs(y(i)) - T)$, thus $x_s(i) + x_p(i) \neq y(i)$.
For coefficients below threshold\camera{,} ${ y(i) }$ can be recovered
trivially since $x_s(i) = 0$ and $x_p(i) = y(i)$.
Note that incorrect sign in the public image occurs only for
coefficients $y(i)$ above threshold, and by definition, for all those
coefficients the public value is $x_p(i)=T$.
Note also that removing these signs increases significantly the
distortion in the public images and makes it more challenging for an
attacker to approximate the original image based on only the public
one.

In summary, the reconstruction can be written as a series of linear
operations:
\begin{eqnarray}
	\label{eq:reconst_noproc}
	 {\bf y} &=& {\bf S_p}\cdot {\bf a_p} + 
        {\bf S_s}\cdot {\bf  a_s} + \left( {\bf S_s} - {\bf S_s}^2
        \right) \cdot {\bf w}
\end{eqnarray} where the first two terms correspond to directly adding
the correspondig blocks from the public and secret images, while the
third term is a correction factor to account for the incorrect sign of
some coefficients in the public image.
This correction factor is based on the sign of the coefficients in the
secret image and distinguishes three cases.
If $x_s(i) = 0$ or $x_s(i)> 0$ then $y(i) = x_s(i) + x_p(i) $ (no
correction), while if $x_s(i)< 0$ we have
\[
y(i) = x_s(i) + x_p(i) - 2T = x_s(i) +T -2T = x_s(i) -T . 
\] Note that the operations can be very easily represented and
implemented with if/then/else conditions, but the algebraic
representation of (\ref{eq:reconst_noproc}) will be needed to
determine how to operate when the public image has been subject to
server-side processing.
In particular, from (\ref{eq:reconst_noproc}), and given that the DCT
is a linear operator, it becomes apparent that it would be possible to
reconstruct the images in the pixel domain.
That is, we could convert ${\bf S_p}\cdot {\bf a_p} $, ${\bf S_s}\cdot
{\bf a_s} $ and $\left( {\bf S_s} - {\bf S_s}^2 \right) \cdot {\bf w}$
into the pixel domain and simply add these three images pixel by
pixel.
Further note that the third image, the correction factor, does not
depend on the public image and can be completely derived from the
secret image.
%



\camera{
}
We now consider the case 
where the PSP applies a linear operator ${\bf A}$ to the public
part. 
Many interesting image transformations such as filtering,
cropping\footnote{\camera{Cropping at 8x8 pixel boundaries is a linear
  operator; cropping at arbitrary boundaries can be approximated by
  cropping at the nearest 8x8 boundary.}
},
scaling (resizing), and overlapping can be expressed by linear
operators.
Thus, when the public part is requested from the PSP, ${\bf A}\cdot
{\bf S_p} \cdot {\bf a_p}$ will be received.
Then the goal is for the recipient to reconstruct ${\bf A}\cdot {\bf
y}$ given the processed public image ${\bf A} \cdot {\bf S_p}\cdot
{\bf a_p}$ and the unprocessed secret information.
Based on the reconstruction formula of  (\ref{eq:reconst_noproc}), and
the linearity of ${\bf A}$, it is clear 
that the desired reconstruction can be obtained as follows
\begin{equation}
	\label{eq:reconst}
	{\bf A} \cdot {\bf y} 
= 
 {\bf A} \cdot {\bf S_p} \cdot {\bf a_p} + {\bf A }
       \cdot  {\bf S_s}\cdot {\bf  a_s} + {\bf A} \cdot \left( {\bf S_s} - {\bf S_s}^2
        \right) \cdot {\bf w}
\end{equation}
Moreover, since the DCT transform is also linear, these operations can
be applied directly in the pixel domain, without needing to find a
transform domain representation. As an example, if cropping is
involved, it would be enough to crop the private image and the image
obtained by applying an inverse DCT to $\left( {\bf S_s} - {\bf S_s}^2
        \right) \cdot {\bf w}$.

\camera{We have left an exploration of nonlinear operators to future work.
  It may be possible to support certain types of non-linear
  operations, such as pixel-wise color remapping, as found in popular
  apps (e.g., Instagram).
%
%
If such operation can be represented as one-to-one mappings for all
legitimate values\footnote{\camera{Often, this is the case for most
color remapping operations.}}, e.g.
0-255 RGB values, we can achieve the same level of reconstruction
quality as the linear operators: at the recipient, we can reverse the
mapping on the public part, combine this with the unprocessed
secret part, and re-apply the color mapping on the resulting image.
However, this approach can result in some loss and we have left a
quantitative exploration of this loss to future work.
} 
%




\subsection{\camera{Algorithmic} Properties of \ppis}

\mypar{Privacy Properties.}
By encrypting significant signal information, \ppis can preserve the
privacy of images by distorting them and by foiling
detection and recognition algorithms (Section~\ref{sec:eval}).
Given only the public part, the attacker can guess the threshold $T$
by assuming it to be the most frequent non-zero value.
If this guess is correct, the attacker knows the positions of the
significant coefficients, but not the range of values of these
coefficients.
Crucially, the sign of the coefficient is also not known.
Sign information tends to be ``random'' in that positive and negative
coefficients are almost equally likely and there is very limited
correlation between signs of different coefficients, both within a
block and across blocks.
It can be shown that if the sign is unknown, and no prior information
exists that would bias our guess, it is actually best, in terms of
mean-square error (MSE), to replace the coefficient with unknown sign
in the public image by 0.\footnote{\camera{If an adversary sees T in the 
public part, replacing it with 0 will have an MSE of $T^2$.
However, if we use any non-zero values as a guess, MSE will be at least
$0.5 \times {( 2T )}^2 = {2T}^2$ because we will have a wrong sign 
with probability 0.5 and we know that the magnitude is at least equal to T.
}}

%
%
%

Finally, we observe that \emph{proving} the privacy properties of our
approach is challenging.
If the public part is leaked from the PSP, proving that no human can
extract visual information from the public part would require having
an accurate understanding of visual perception.
%
Instead, we rely on metrics
commonly used in the signal processing community
\camera{in our evaluation (Section~\ref{sec:eval})}.
We note that the prevailing methodology in the signal processing
community for evaluating the efficacy of image and video privacy is
empirical subjective evaluation using user studies, or objective
evaluation using metrics~\camera{\cite{richardson2011h}}.
In Section~\ref{sec:eval}, we resort to an objective metrics-based
evaluation, showing the performance of \ppis on several image corpora.

\mypar{Other Properties.} 
\ppis satisfies the other requirements we have discussed above.
It leaves, in the clear, a JPEG-compliant image (the public part), on
which the PSP can perform transformations to save storage and
bandwidth.
The threshold $T$ permits trading off increased storage for increased
privacy; for images whose signal content is in the DC component and a
few highly-valued coefficients, the secret part can encode most of
this content, while the public part contains a significant fraction of
the volume of the image in bytes.
As we show in our evaluation later, most images are sparse and satisfy
this property.
Finally, our approach of encoding the large coefficients decreases the
entropy both in the public and secret parts, resulting in better
compressibility and only slightly increased overhead overall relative
to the unencrypted compressed image.

However, the \ppis algorithm has an interesting consequence: since the
secret part cannot be scaled (because, in general, the transformations
that a PSP performs cannot be known a priori) and must be downloaded
in its entirety, the bandwidth savings from \ppis will always be lower
than downloading a resized original image.
The size of the secret part is determined by $T$: higher values of $T$
result in smaller secret parts, but provide less privacy, a trade-off
we quantify in Section~\ref{sec:eval}.

\section{\ppis: System Design}
\label{sec:sysarch}

\begin{figure}
    \centering
    \includegraphics[viewport=20 180 700 540, scale=0.32, clip=true]{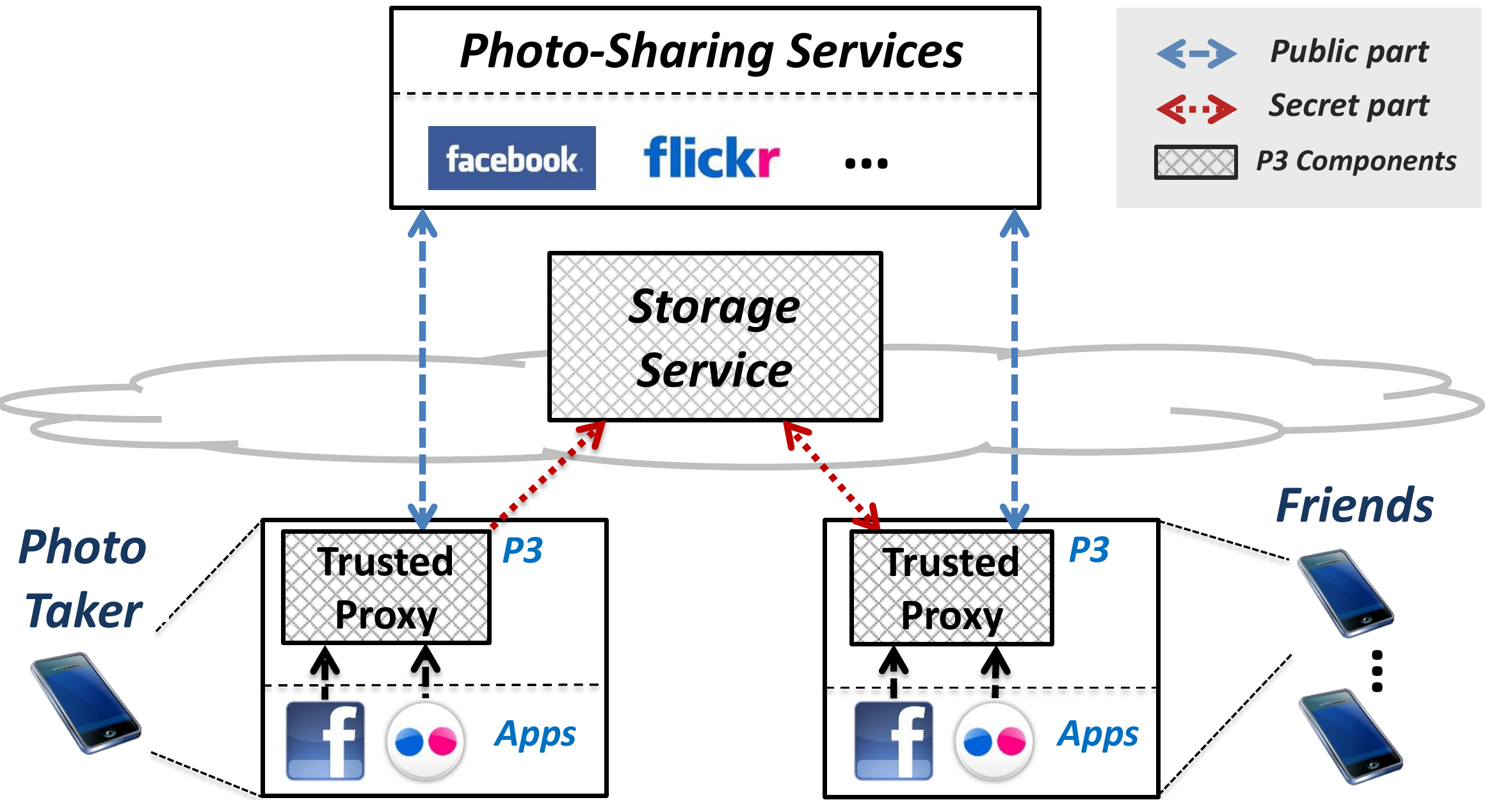}
    \caption{\ppis System Architecture}
    \label{fig:sysarch}
    \vspace{-2ex}
\end{figure}

In this section, we describe the design of a system for privacy
preserving photo sharing system.
This system, also called \ppis, has two desirable properties described
earlier.
First, it requires no software modifications at the PSP.
Second, it requires no modifications to client-side browsers or image
management applications, and only requires a small footprint software
installation on clients.
%
These properties permit \camera{fairly easy} deployment of privacy-preserving
photo sharing.

\subsection{\ppis Architecture and Operation}
Before designing our system, we explored the protocols used by PSPs
for uploading and downloading photos.
Most PSPs use HTTP or HTTPS to upload messages; we have verified this
for Facebook, Picasa Web, Flickr, PhotoBucket, Smugmug, and
Imageshack.
This suggests a relatively simple interposition architecture, depicted
in Figure~\ref{fig:sysarch}.
In this architecture, browsers and applications are configured to use
a local HTTP/HTTPS \emph{proxy} and all accesses to PSPs go through
the proxy.
The proxy manipulates the data stream to achieve privacy preserving
photo storage, in a manner that is transparent both to the PSP and the
client.
In the following paragraphs, we describe the actions performed by the
proxy at the sender side and at one or more recipients.


\mypar{Sender-side Operation.} 
When a sender transmits the photo taken by built-in camera, the local
proxy acts as a middlebox and splits the uploaded image into a public
and a secret part (as discussed in Section~\ref{sec:approach}).
Since the proxy resides on the client device (and hence is within the
trust boundary per our assumptions, Section~\ref{sec:motiv}), it is
reasonable to assume that the proxy can decrypt and encrypt HTTPS
sessions in order to encrypt the photo.

We have not yet discussed how photos are encrypted; in our current
implementation, we assume the existence of a symmetric shared key
between a sender and one or more recipients.
This symmetric key is assumed to be distributed 
out of band.

Ideally, it would have been preferable to store both the public and
the secret parts on the PSP.
Since the public part is a JPEG-compliant image, we explored methods
to embed the secret part within the public part.
The JPEG standard allows users to embed arbitrary application-specific
\emph{markers} with application-specific data in images; the standard
defines 16 such markers.
We attempted to use an application-specific marker to embed the secret
part; unfortunately, at least 2 PSPs (Facebook and Flickr) strip all
application-specific markers.

Our current design therefore stores the secret part on a cloud storage
provider (in our case, Dropbox).
Note that because the secret part is encrypted, we do not assume that
the storage provider is trusted.
%

Finally, we discuss how photos are named.
When a user uploads a photo to a PSP, that PSP may transform the photo
in ways discussed below.
Despite this, most photo-sharing services (Facebook, Picasa Web,
Flickr, Smugmug, and Imageshack\footnote{PhotoBucket does not, which
  explains its vulnerability to fusking, as discussed earlier}) assign
a unique ID for all variants of the photo.
This ID is returned to the client, as part of the
API~\cite{FacebookAPI, FlickrAPI}, when the photo is updated.

\ppis's sender side proxy performs the following operations on the
public and secret parts.
First, it uploads the public part to the PSP either using HTTP or
HTTPS (e.g., Facebook works only with HTTPS, but Flickr supports
HTTP).
This returns an ID, which is then used to name a file containing the
secret part.
This file is then uploaded to the storage provider.

\mypar{Recipient-side Operation.}
Recipients are also configured to run a local web proxy.
A client device downloads a photo from a PSP using an HTTP get
request. 
The URL for the HTTP request contains the ID of the photo
being downloaded.
When the proxy sees this HTTP request, it passes \camera{the} request on to the
PSP, but also initiates a concurrent download of the secret part from
the storage provider using the ID embedded in the URL.
When both the public and secret parts have been received, the proxy
performs the decryption and reconstruction procedure discussed in
Section~\ref{sec:approach} and passes the resulting image to the
application as the response to the HTTP get request.
However, note that a secret part may be reused multiple times: for
example, a user may first view a thumbnail image and then download a
larger image.
In these scenarios, it suffices to download the secret part once so
the proxy can maintain a cache of downloaded secret parts in order to
reduce bandwidth and improve latency.

There is an interesting subtlety in the photo reconstruction process.
As discussed in Section~\ref{sec:approach}, when the server-side
transformations are known, nearly exact reconstruction is
possible\footnote{The only errors that can arise are due to storing
  the correction term in Section~\ref{sec:approach} in a lossy JPEG
  format that has to be decoded for processing in the pixel
  domain. Even if quantization is very fine, errors maybe introduced
  because the DCT transform is real valued and pixel values are
  integer, so the inverse transform of $\left( {\bf S_s} - {\bf S_s}^2
  \right) {\bf w}$ will have to be rounded to the nearest integer
  pixel value.}.
In our case, the precise transformations are not known, in general, to
the proxy, so the problem becomes more challenging.

By uploading photos, and inspecting the results, we are able to tell,
generally speaking, what kinds of transformations PSPs perform.
For instance, Facebook transforms a baseline JPEG image to a
progressive format and at the same time wipes out all irrelevant markers.
Both Facebook and Flickr statically resize the uploaded image with
different sizes; for example, Facebook generates at least three files
with different resolutions, while Flickr generates a series of
fixed-resolution images whose number depends on the size of the
uploaded image.
We cannot tell if these PSPs actually store the original images or
not, and we conjecture that the resizing serves to limit storage and
is also perhaps optimized for common case devices.
For example, the largest resolution photos stored by Facebook is
720x720, regardless of the original resolution of the image.
In addition, Facebook can dynamically resize and crop an image; the
cropping geometry and the size specified for resizing are both
encoded in the HTTP get URL, so the proxy is able to determine those
parameters.
Furthermore, by inspecting the JPEG header, we can tell some kinds of
transformations that may have been performed: e.g., whether baseline
image was converted to progressive or vice a versa, what sampling
factors, cropping and scaling etc. were applied.

However, some other critical image processing parameters are not
visible to the outside world.
For example, the process of resizing an image using down sampling is
often accompanied by a filtering step for antialiasing and may be
followed by a sharpening step, together with a color adjustment step
on the downsampled image.
Not knowing which of these steps have been performed, and not knowing
the parameters used in these operations, the reconstruction procedure
can result in lower quality images.

To understand what transformations have been performed, we are reduced
to searching the space of possible transformations for an outcome that
matches the output of transformations performed by the PSP\footnote{
  This approach is clearly fragile, since the PSP can change the kinds
  of transformations they perform on photos. Please see the discussion
  below on this issue.
}.
Note that this reverse engineering need only be done when a PSP
re-jiggers its image transformation pipeline, so it should not be too
onerous.
Fortunately, for Facebook and Flickr, we were able to get reasonable
reconstruction results on both systems (Section~\ref{sec:eval}).
These reconstruction results were obtained by exhaustively searching
the parameter space with salient options based on commonly-used
resizing techniques~\cite{ResizeTechnique}.
More precisely, we select several candidate settings for colorspace
conversion, filtering, sharpening, enhancing, and gamma corrections,
and then compare the output of these with that produced by the PSP.
Our reconstruction results are presented in Section~\ref{sec:eval}.
%



%

\subsection{Discussion}
\label{sec:sysarch-discuss}
%
%
\mypar{Privacy Properties.} 
Beyond the privacy properties of the \ppis algorithm, the \ppis system
achieves the privacy goals outlined in Section~\ref{sec:motiv}.
Since the proxy runs on the client for both sender and receiver, the
trusted computing base for \ppis includes the software and hardware
device on the client.
It may be possible to reduce the footprint of the trusted computing
base even further using a trusted platform module~\cite{TPM} and
trusted sensors~\cite{TrustedSensor}, but we have deferred that to future
work.

\ppis's privacy depends upon the strength of the symmetric key used
to encrypt in the secret part. 
We assume the use of AES-based symmetric keys, distributed out of band.
%
%
Furthermore, as discussed above, in \ppis the storage provider cannot
leak photo privacy because the secret part is encrypted.
The storage provider, or for that matter the PSP, can tamper with
images and hinder reconstruction; protecting against such tampering is
beyond the scope of the paper.
For the same reason, eavesdroppers can similarly potentially tamper
with the public or the secret part, but cannot leak photo privacy.



\para{PSP Co-operation.} 
The \ppis design we have described assumes no co-operation from the
PSP.
As a result, this implementation is fragile and a PSP can prevent
users from using their infrastructure to store \ppis's public parts.
For instance, they can introduce complex nonlinear transformations on
images in order to foil reconstruction.
They may also run simple algorithms to detect images where
coefficients might have been thresholded, and refuse to store such
images.
%

Our design is merely a proof of concept that the technology exists to
transparently protect the privacy of photos,  without requiring
infrastructure changes or significant client-side modification.
Ultimately, PSPs will need to cooperate in order for photo privacy to
be possible, and this cooperation depends upon the implications of
photo sharing on their respective business models.

At one extreme, if only a relatively small fraction of a PSP's user base
uses \ppis, a PSP may choose to benevolently ignore this use (because
preventing it would require commitment of resources to reprogram their
infrastructure).
At the other end, if PSPs see a potential loss in revenue from not
being able to recognize objects/faces in photos, they may choose to
react in one of two ways: shut down \ppis, or offer photo privacy for
a fee to users.
However, in this scenario, a significant number of users see value
in photo privacy, so we believe that PSPs will be incentivized to
offer privacy-preserving storage for a fee.
In a competitive marketplace, even if one PSP were to offer
privacy-preserving storage as a service, others will likely follow
suit.
For example, Flickr  already has a ``freemium'' business model and can
simply offer privacy preserving storage to its premium subscribers.

If a PSP were to offer privacy-preserving photo storage as a service,
we believe it will have incentives to use a \ppis like approach (which
permits image scaling and transformations), rather than end to end
encryption.
With \ppis, a PSP can assure its users that it is only able to see the
public part (reconstruction would still happen at the client), yet
provide (as a service) the image transformations that can reduce
user-perceived latency (which is an important consideration
for retaining users of online services~\camera{\cite{Haystack}}).
%

Finally, with PSP co-operation, two aspects of our \ppis design
become simpler.
First, the PSP image transformation parameters would be known, so
higher quality images would result.
Second, the secret part of the image could be embedded within the
public part, obviating the need for a separate online storage
provider.

\mypar{Extensions.}
Extending this idea to video is feasible, but left for future work.
As an initial step, it is possible to introduce the privacy preserving
techniques only to the I-frames, which are coded independently using
tools similar to those used in JPEG.
Because other frames in a ``group of pictures'' are coded using an
I-frame as a predictor, quality reductions in an I-frame propagate
through the remaining frames.
In future work, we plan to study video-specific aspects, such as how
to process motion vectors or how to enable reconstruction from a
processed version of a public video.
%

%
%



\section{Evaluation}
\label{sec:eval}

In this section, we report on an evaluation of \ppis.
Our evaluation uses objective metrics to characterize the privacy
preservation capability of \ppis, and it also reports, using a
full-fledged implementation, on the processing overhead induced by
sender and receiver side encryption.

\subsection{Methodology}

\mypar{Metrics.}
Our first metric for \ppis performance is the \emph{storage overhead}
imposed by selective encryption.
Photo storage space is an important consideration for PSPs, and a
practical scheme for privacy preserving photo storage must not incur
large storage overheads.
\iftechrep
We then measure the efficacy of privacy preservation using PSNR (peak
signal-to-noise ratio), a metric commonly used in signal
processing. While the shortcomings of this metric in terms of quantifying
perceptual quality are well known, it does provide a simple objective
way of quantifying degradation. Note also that the public images 
\camera{that are highly degraded with values of PSNR} will be commonly
agreed to represent very poor quality. 
To complement PSNR, we also present the visual representation of the
public part of an image, to let the reader judge the efficacy of
\ppis; lack of space prevents us from a more detailed exposition.
\fi
We then evaluate the efficacy of privacy preservation by measuring the
performance of state-of-the-art edge and face detection algorithms,
\camera{
the SIFT feature extraction algorithm, and a face recognition algorithm 
}
on \ppis.
We conclude the evaluation of privacy by discussing the efficacy of
guessing attacks.
\iftechrep
\fi
Finally, we quantify the reconstruction performance, bandwidth savings
and the processing overhead of \ppis.

\mypar{Datasets.}
We evaluate \ppis using \camera{four} image datasets.
First, as a baseline, we use the ``miscellaneous'' volume in the
USC-SIPI image dataset~\cite{USCdataset}.
This volume has 44 color and black-and-white images and contains
various objects, people, scenery, and so forth, and contains many
canonical images (including Lena) commonly used in the image
processing community.
Our second data set is from INRIA~\cite{INRIAdataset}, and contains 1491
full color images from vacation scenes including a mountain, a river,
a small town, other interesting topographies, etc.
This dataset contains has greater diversity than the USC-SIPI dataset
in terms of both resolutions and textures; its images vary in size up
to 5 MB, while the USC-SIPI dataset's images are all under 
\camera{1 MB.}
%
%

We also use the Caltech face dataset~\cite{Caltechdataset} for our
face detection experiment.
This has 450 frontal color face images of about 27 unique faces
depicted under different circumstances (illumination, background,
facial expressions, etc.).
All images contain at least one large dominant face, and zero or more
additional faces.
\camera{Finally, the Color FERET Database~\cite{FERET} is used for
our face recognition experiment.
This dataset is specifically designed for developing, testing, and
evaluating face recognition algorithms, and contains 11,338 facial
images, using 994 subjects at various angles.}

\mypar{Implementation.}
We also report results from an implementation for
Facebook~\cite{Facebook}.
We chose the Android 4.x mobile operating system as our client
platform, since the bandwidth limitations together with the
availability of camera sensors on mobile devices motivate our work.
The \emph{mitmproxy} software tool~\cite{MITMPROXY} is used as a
trusted man-in-the-middle proxy entity in the system.
To execute a mitmproxy tool on Android, we used the
\textit{kivy/python-for-android} software~\cite{KivyP4A}.
Our algorithm described in Section~\ref{sec:approach} is implemented
based on the code maintained by the Independent JPEG Group, version
8d~\cite{IJG}.
We report on experiments conducted by running this prototype on
Samsung Galaxy S3 smartphones.

\begin{figure}[t]
    \centering
    \includegraphics[viewport=0 150 720 540, scale=0.24, clip=true]{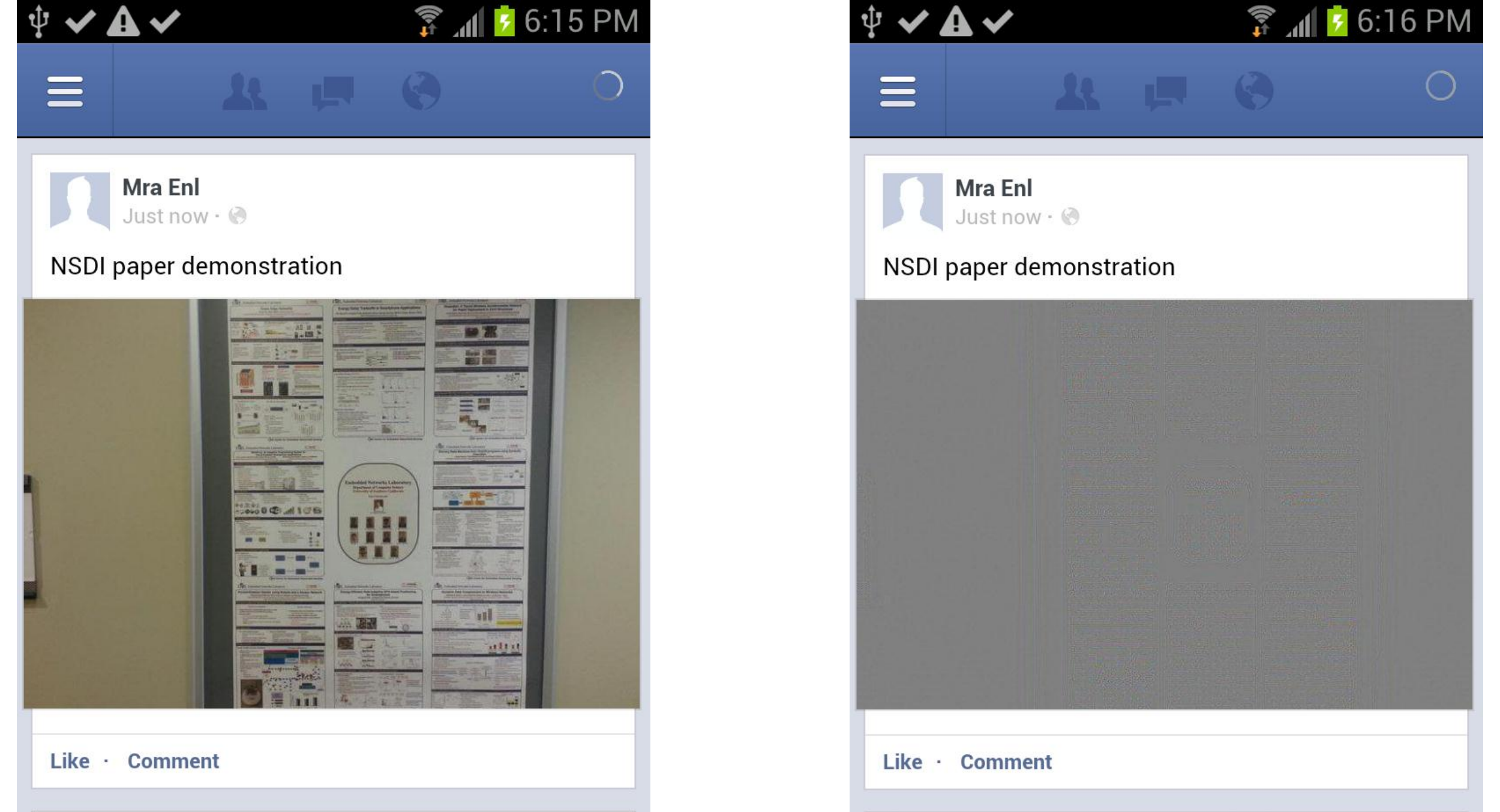}
    \caption{Screenshot(Facebook) with/without decryption}
    \label{fig:screen}
\end{figure}

Figure~\ref{fig:screen} shows two screenshots of a Facebook page, with
two photos posted.
The one on the left is the view seen by a mobile device which has our
recipient-side decryption and reconstruction algorithm enabled.
On the right is the same page, without that algorithm (so only the
public parts of the images are visible).

%
%


%
%

\subsection{Evaluation Results}

In this section, we first report on the trade-off between the
threshold parameter and storage size in \ppis.
We then evaluate various privacy metrics, and conclude with an
evaluation of reconstruction performance, bandwidth, and processing
overhead.

\subsubsection{The Threshold vs. Storage Tradoff}

\begin{figure}[t]
	\centering
	\subfigure[USC-SIPI]{
		\includegraphics[viewport=0 0 275 270, width=1.5in, clip=true]{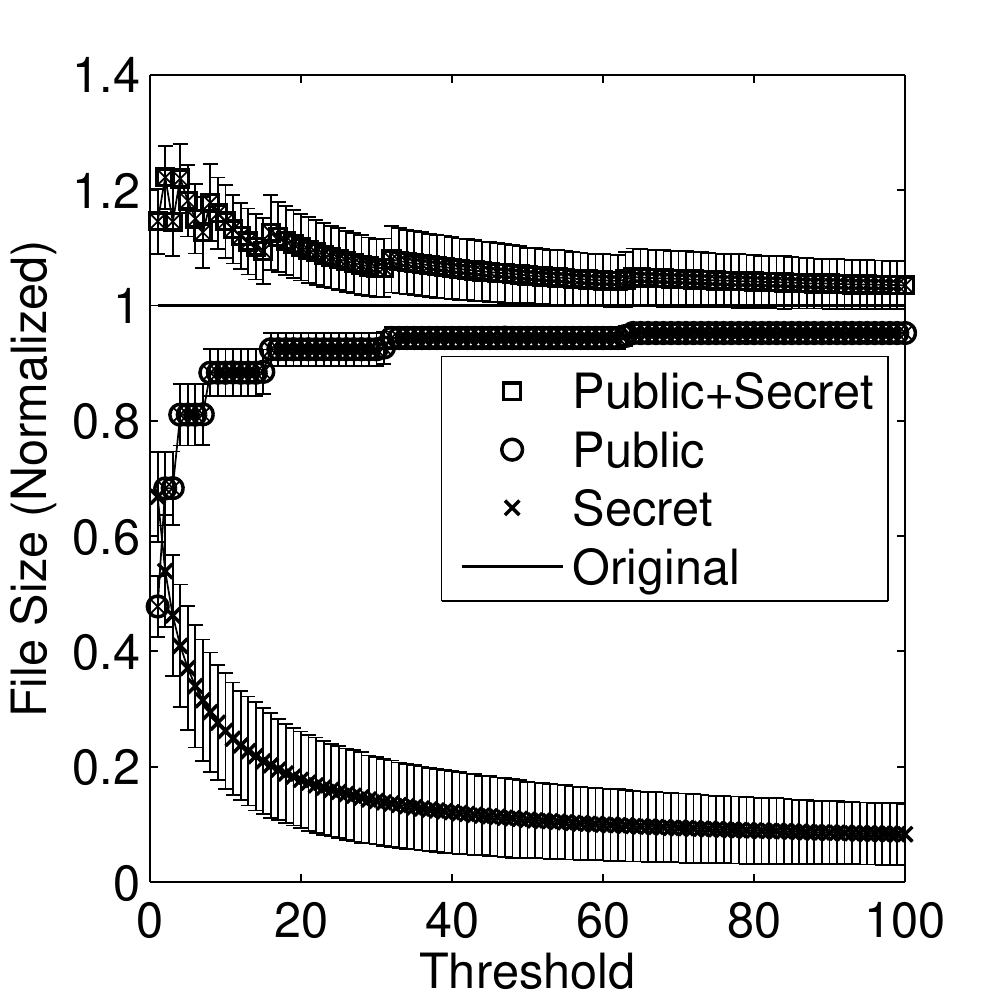}
		\label{fig:algo_size_tradeoff_usc}
	}
	\hspace{-2ex}
	\subfigure[INRIA]{
		\includegraphics[viewport=0 0 275 270, width=1.5in, clip=true]{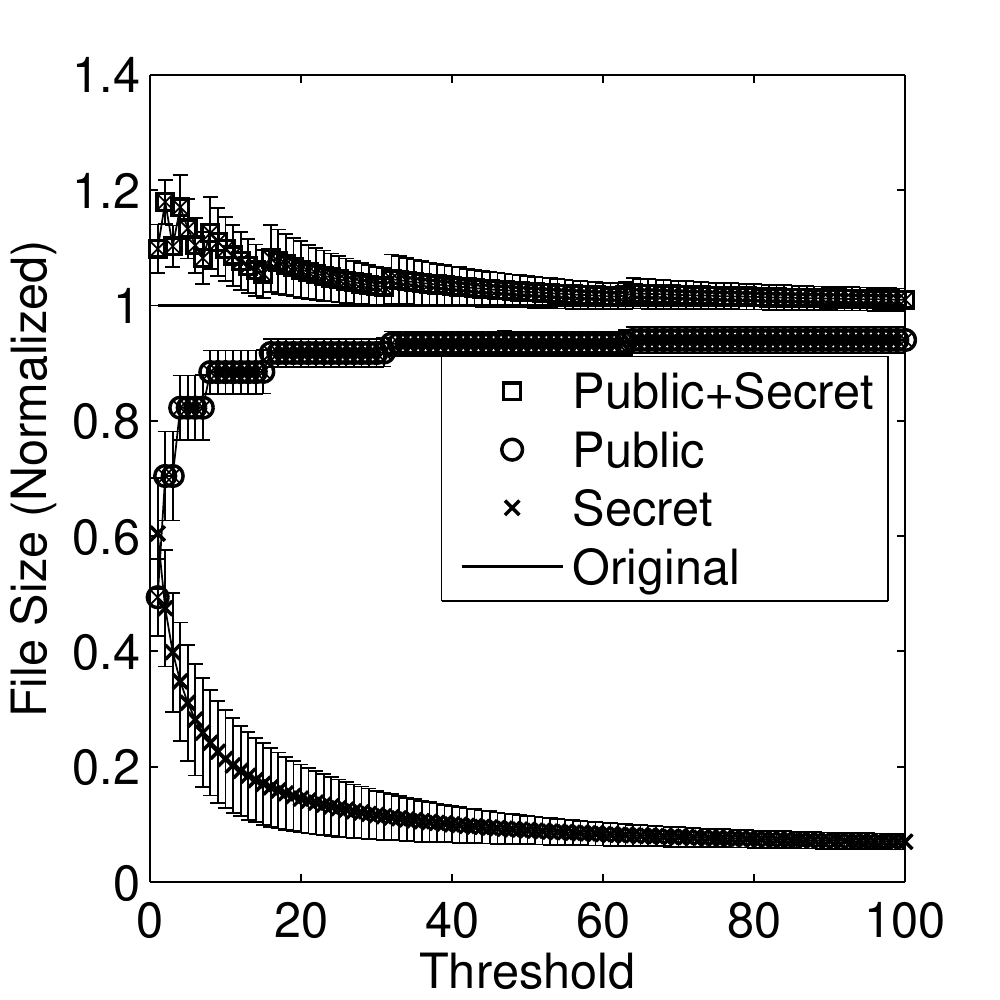}
		\label{fig:algo_size_tradeoff_inria}
	}
    \vspace{-2ex}
	\caption{Threshold vs. Size \camera{(error bars=stdev)}}
	\label{fig:algo_size_tradeoff}
\end{figure}

\iftechrep
\begin{figure}[t]
	\centering
	\subfigure[USC-SIPI]{
		\includegraphics[viewport=0 0 275 205, width=1.55in, clip=true]{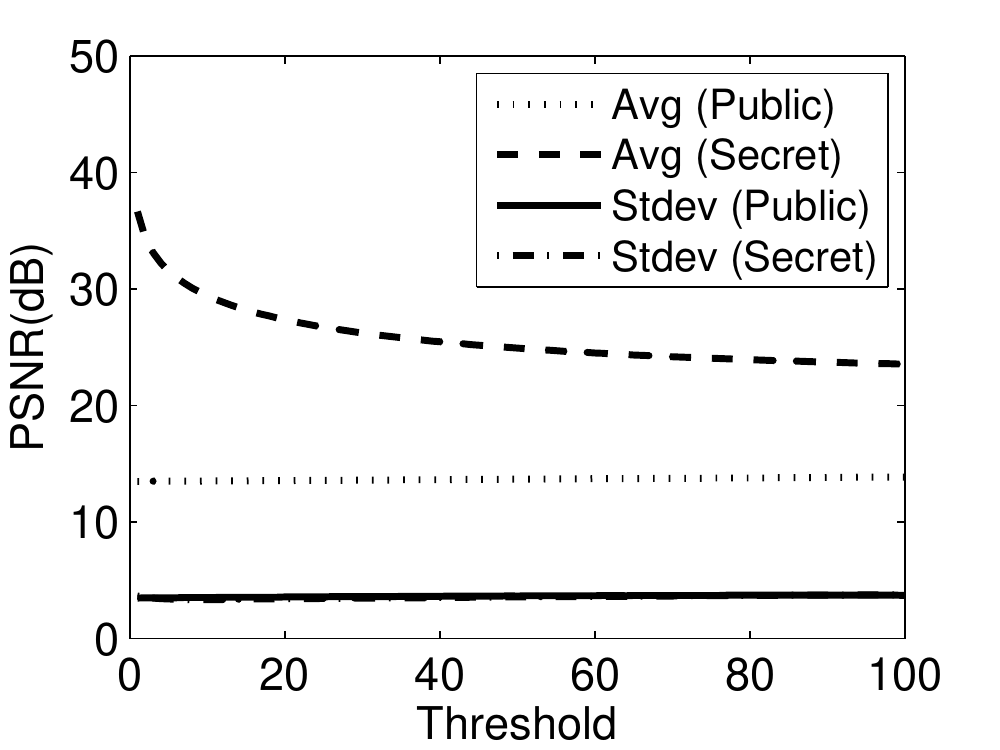}
		\label{fig:algo_psnr_usc}
	}
	\hspace{-3ex}
	\subfigure[INRIA]{
		\includegraphics[viewport=0 0 275 205, width=1.55in, clip=true]{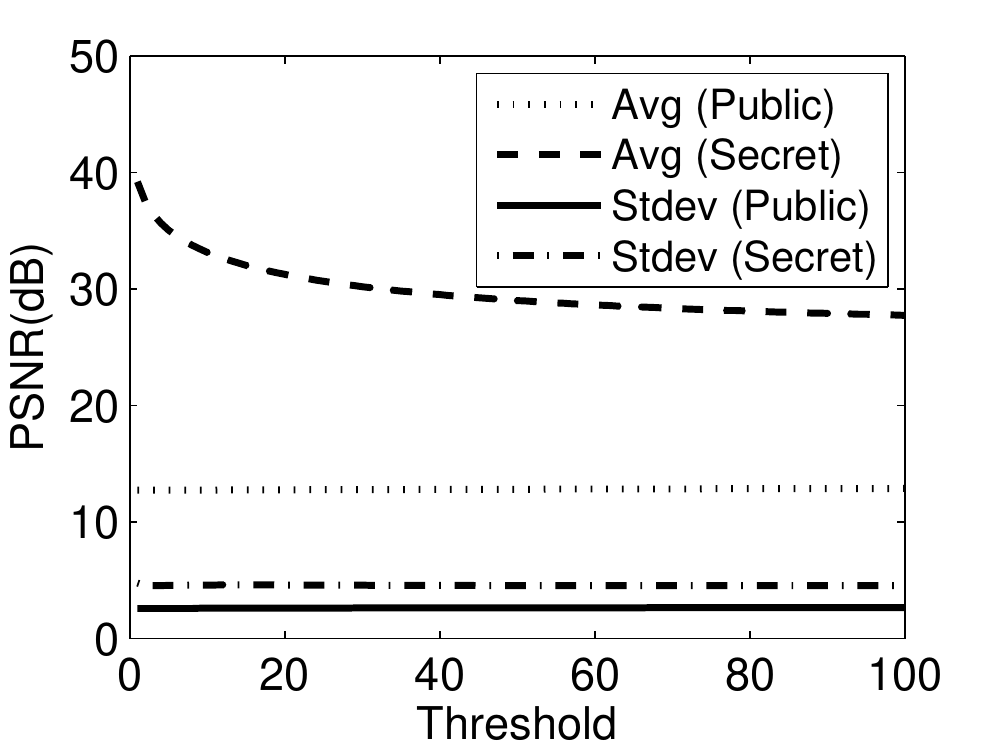}
		\label{fig:algo_psnr_inria}
	}
	\vspace{-3ex}
	\caption{PSNR results}
	\label{fig:algo_psnr}
\end{figure}
\fi

\begin{figure}[t]
    \centering
    \subfigure[Public Part]{
        \includegraphics[width=3.1in, clip=true]{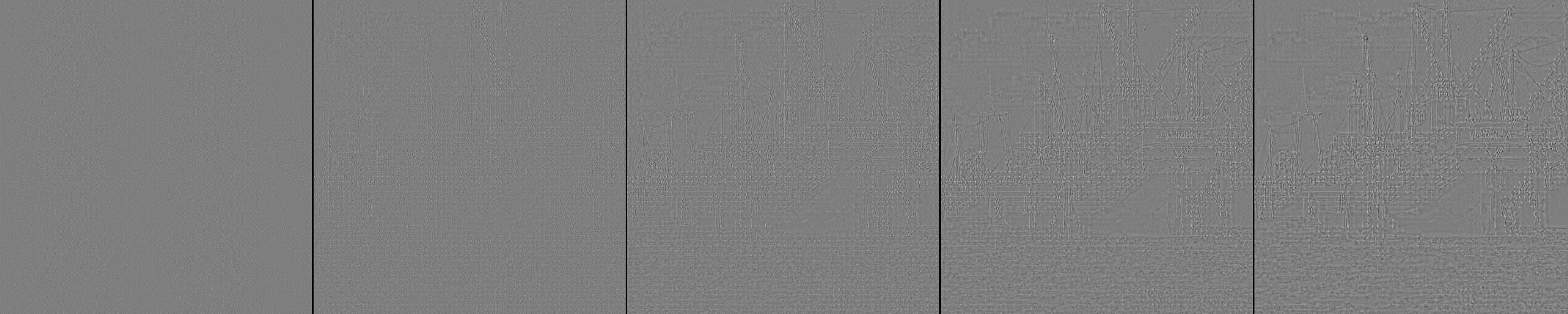}
        \label{fig:algo_base_public_boat}
    }
    \subfigure[Secret Part]{
        \includegraphics[width=3.1in, clip=true]{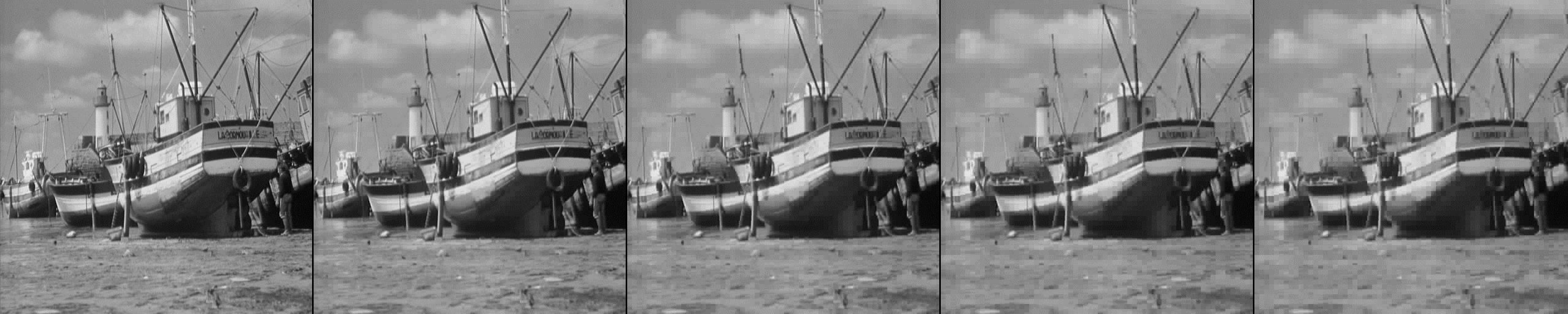}
        \label{fig:algo_base_secret_boat}
    }
    \vspace{-3ex}
    \caption{Baseline - Encryption Result (T: 1,5,10,15,20)}
    \label{fig:algo_base}
    \vspace{-3ex}
\end{figure}

In \ppis, the threshold $T$ is a tunable parameter that trades off
storage space for privacy: at higher thresholds, fewer coefficients
are in the secret part but more information is exposed in the public
part.
Figure~\ref{fig:algo_size_tradeoff} reports on the size of the public
part (a JPEG image), the secret part (an encrypted JPEG image), and
the combined size of the two parts, as a fraction of the size of the
original image,  for different threshold values $T$.
One interesting feature of this figure is that, despite the
differences in size and composition of the two data sets, their size
\emph{distribution as a function of thresholds is qualitatively
  similar}.
At low thresholds (near 1), the combined image sizes exceed the
original image size by about 20\%, with the public and secret parts
being each about 50\% of the total size.
While this setting provides excellent privacy, the large size of the
secret part can impact bandwidth savings; recall that, in \ppis, the
secret part has to be downloaded in its entirety even when the public
part has been resized significantly.
Thus, it is important to select a better operating point where the
size of the secret part is smaller.

Fortunately, the shape of the curve of
Figure~\ref{fig:algo_size_tradeoff} for \emph{both datasets} suggests
operating at the knee \camera{of the ``secret'' line} (at a threshold
of in the range of 15-20), where the secret part is about 20\% of the original
image, and the \emph{total storage overhead is about 5-10\%}.
Figure~\ref{fig:algo_base}, which depicts the public and secret parts
(recall that the secret part is also a JPEG image) of a canonical
image from the USC-SIPI dataset, shows that for thresholds in this
range \camera{minimal} visual information is present in the public part,
with all of it being stored in the secret part.
We include these images to give readers a visual sense of the efficacy
of \ppis; we conduct more detailed privacy evaluations below.
This suggests that a threshold between 10-20 might provide a good
balance between privacy and storage.
We solidify this finding below.



\subsubsection{Privacy}

\iftechrep
\para{PSNR.} 
One of the earliest objective metrics used for evaluating the quality of
image reconstruction is the peak signal-to-noise ratio (PSNR).
%
In Figure~\ref{fig:algo_psnr}, we present average PSNRs
\camera{and standard deviations of the public and secret 
part of the USC-SIPI and the INRIA dataset,}
as a function of different thresholds,
when compared to the original image. 

\camera{The secret parts show high PSNRs, especially 
when we consider the fact that 35-40dB is regarded as perceptually 
loseless in the image processing community.
Nonetheless, note that our encryption algorithm uses a single threshold 
across entire image blocks and does not consider block energy distributions.
As a result, even if we get about 40dB in the secret part, 
we can identify non-trivial block effects when we closely observe the image 
(Figure~\ref{fig:algo_base}).
} 
%
It is encouraging that the PSNR values 
\camera{of the public part} are all around 
\camera{10-15 dB,}
and that they increase only slightly with threshold.
The extraction of the DC component into the secret part plays a major
part in leading to such low PSNR values.
%
%
%
%
For the range of (low) PSNRs that we observe here (e.g., around
\camera{15} dB) it is widely accepted that quality is so degraded that
these images are practically useless.
However, this alone is not an indication that \ppis preserves privacy;
an examination of the public part of threshold 100 (not shown) reveals
some of the features in the original image.
At lower thresholds these features are no longer visible
(Figure~\ref{fig:algo_base}), but the difference in PSNR between a
threshold of 10 and 100 is negligible.

For this reason, we consider using several other metrics to quantify
the privacy obtained with \ppis.
These metrics quantify the efficacy of automated algorithms on the
public part; \emph{each automated algorithm can be considered to be
  mounting a privacy attack on the public part.} 
\fi


\begin{figure*}[t]
  \subfigure[Edge Detection]{
      \centering
      \includegraphics[width=1.65in, clip=true]{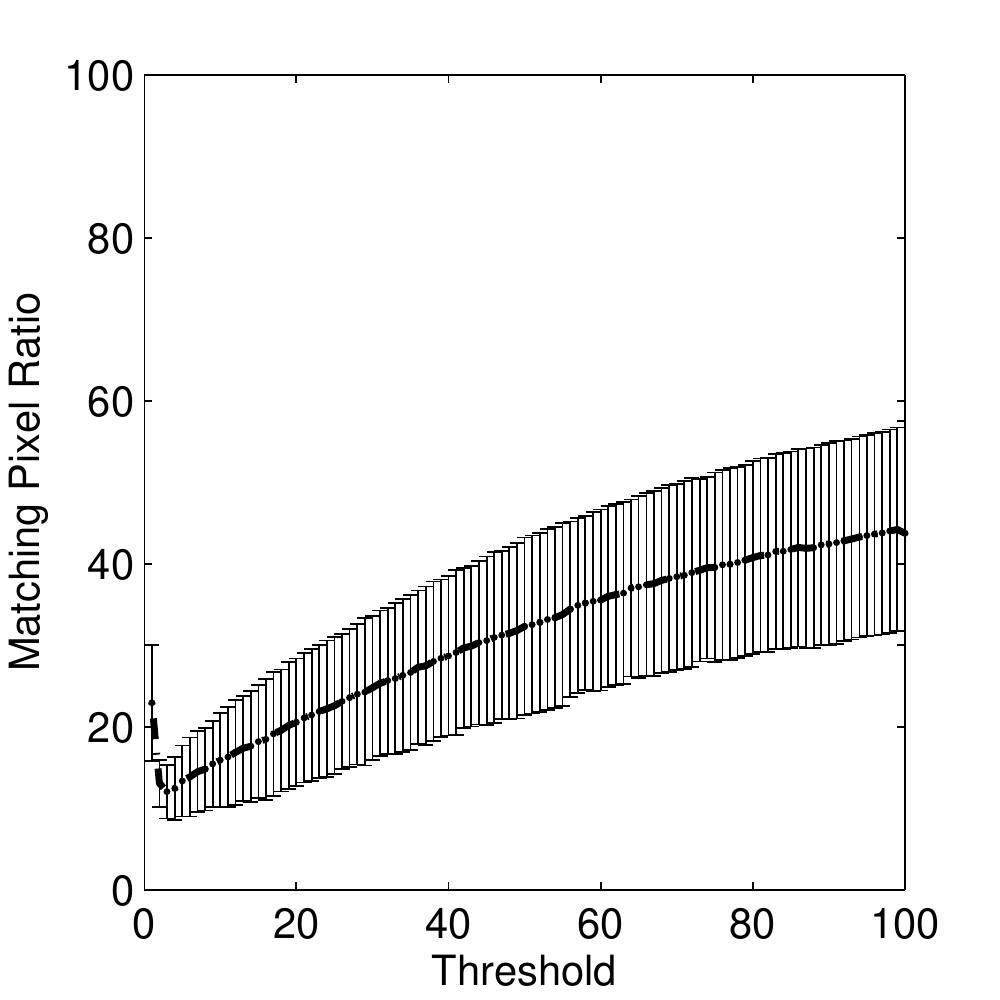}
      \label{fig:algo_canny_hamming}
  }
  \hspace{-3ex}
  \subfigure[Face Detection]{
      \centering
      \includegraphics[width=1.65in, clip=true]{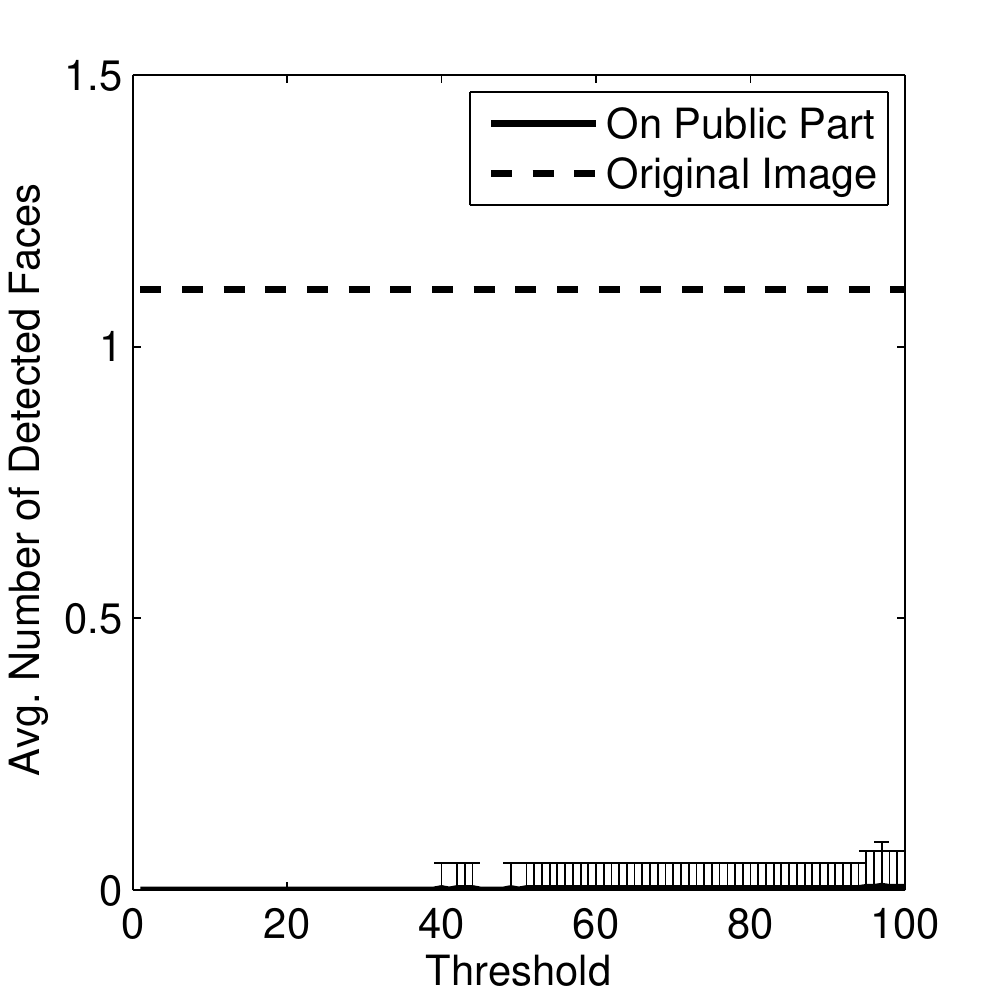}
      \label{fig:algo_facedetect_tradeoff_caltechface}
  }
  \hspace{-3ex}
  \subfigure[SIFT Feature]{
      \centering
      \includegraphics[width=1.65in, clip=true]{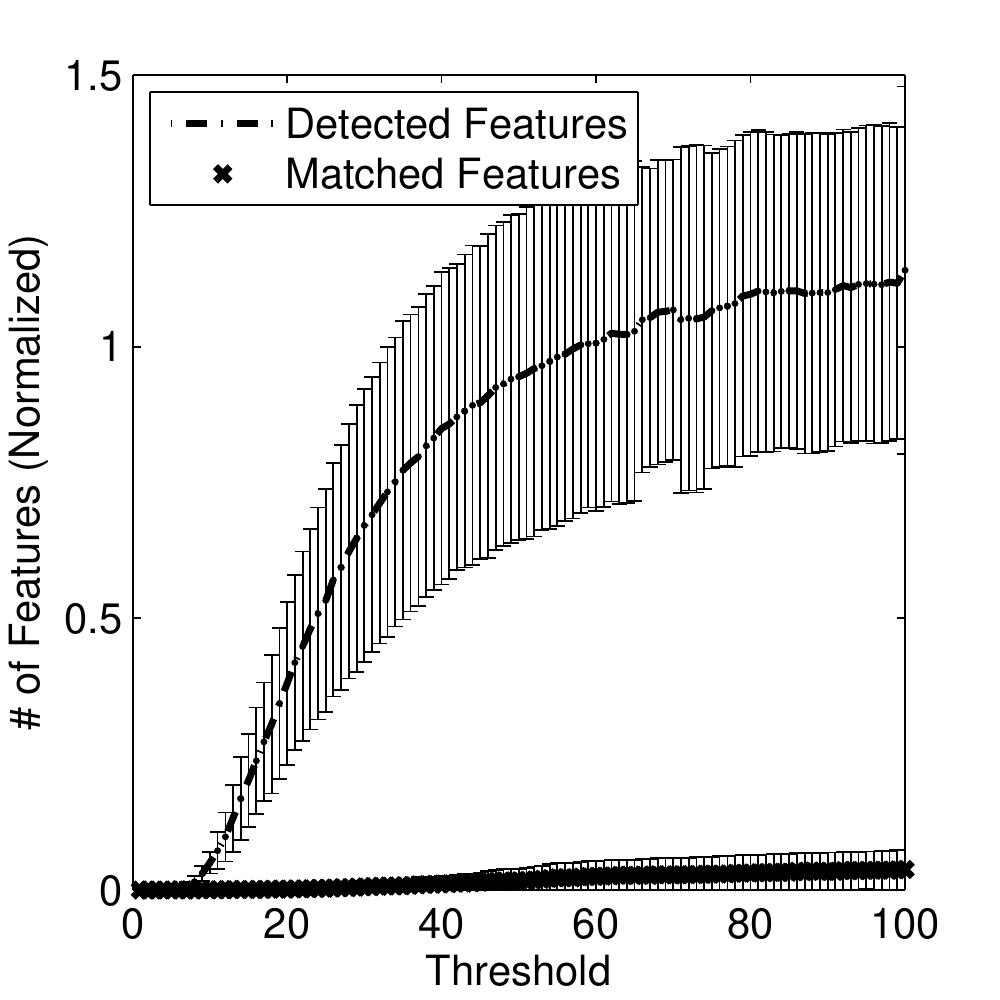}
      \label{fig:algo_sift_tradeoff_usc}
  }
  \hspace{-3ex}
  \subfigure[Face Recognition]{
      \centering
      \includegraphics[width=1.65in, clip=true]{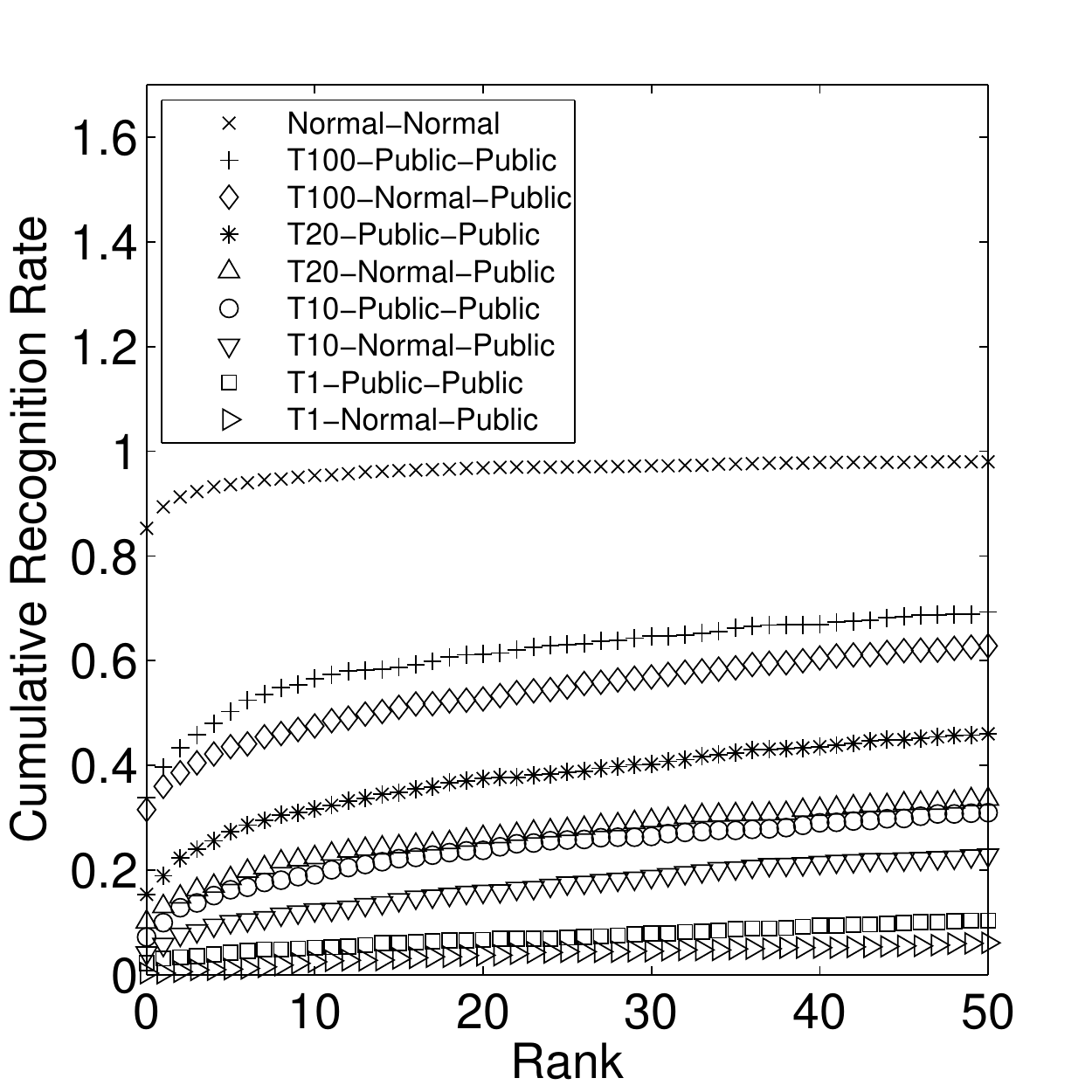}
      \label{fig:algo_facerec_feret_fafb}
  }
  \caption{Privacy on Detection and Recognition Algorithms}
\end{figure*}

\begin{figure}[t]
	\centering
	\subfigure[T=1]{
		\includegraphics[width=3.1in, clip=true]{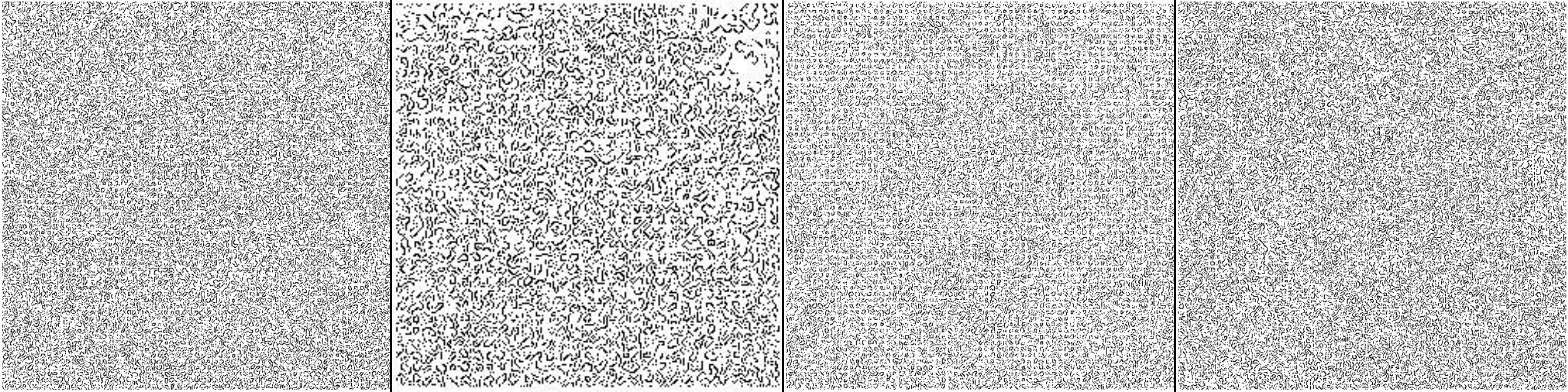}
	}
	\subfigure[T=20]{
		\includegraphics[width=3.1in, clip=true]{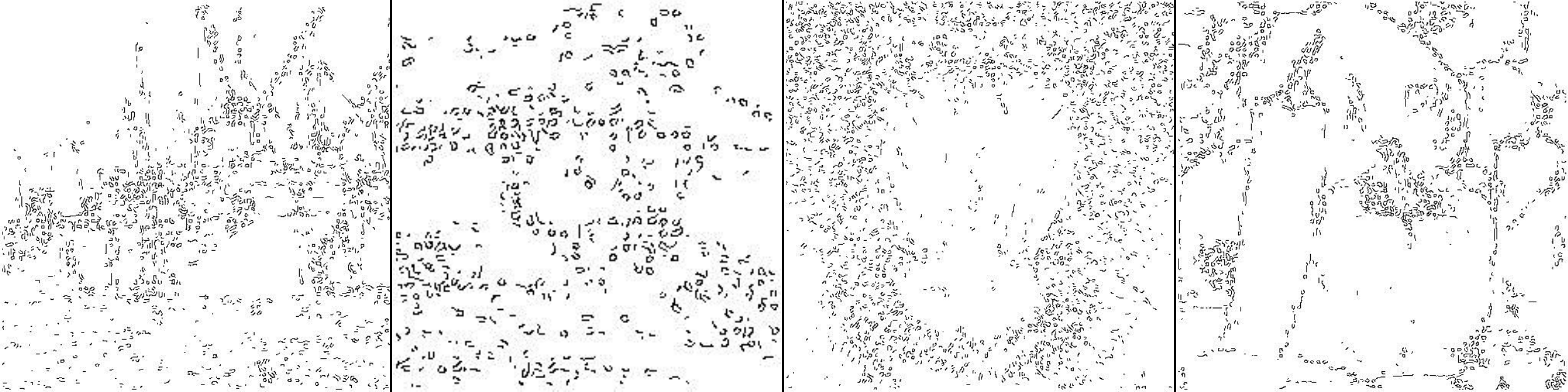}
	}
	\caption{Canny Edge Detection on Public Part}
	\label{fig:algo_edge_public_all}
\end{figure}

\para{Edge Detection.}
Edge detection is an elemental processing step in many signal
processing and machine vision applications, and attempts to discover
discontinuities in various image characteristics.
We apply the well-known Canny edge detector~\cite{CANNY} and its
implementation~\cite{CANNYimpl} to the public part of images in the
USC-SIPI dataset, and present images with the recognized edges in
Figure~\ref{fig:algo_edge_public_all}.
For space reasons, we only show edges detected on the public part of 4
canonical images for a threshold of \camera{1 and} 20. 
%
\camera{The images with a threshold 20} do reveal several ``features'', and signal processing
researchers, when told that these are canonical images from a widely
used data set, can probably recognize these images.
However, a layperson who has not seen the image before very likely
will not be able to recognize any of the objects in the images (the
interested reader can browse the USC-SIPI dataset online to find the
originals).
We include these images to point out that visual privacy is a highly
subjective notion, and depends upon the beholder's prior experiences.
If true privacy is desired, end-to-end encryption must be used.
\ppis provides ``pretty good'' privacy together with the convenience
and performance offered by photo sharing services.

It is also possible to quantify the privacy offered by \ppis for edge
detection attacks.
Figure~\ref{fig:algo_canny_hamming} plots the fraction of matching
pixels in the image obtained by running edge detection on the public
part, and that obtained by running edge detection on the original
image (the result of edge detection is an image with binary pixel
values).
At threshold values below 20, \emph{barely 20\% of the pixels match}; at very
low thresholds, running edge detection on the public part results in a
picture resembling white noise, so we believe the higher matching rate
\camera{shown at low thresholds}
simply results from spurious matches.
We conclude that, for the range of parameters we consider, \ppis is
very robust to edge detection.


\mypar{Face Detection.}
Face detection algorithms detect human faces in photos, and were
available as part of Facebook's face recognition
API, until Facebook shut down the API~\cite{FacebookFace}.
%
%
To quantify the performance of face detection on \ppis, we use the
Haar face detector from the OpenCV library~\cite{OpenCV}, and apply it
to the public part of images from Caltech's face
dataset~\cite{Caltechdataset}.
The efficacy of face detection, as a function of different thresholds,
is shown in Figure~\ref{fig:algo_facedetect_tradeoff_caltechface}.
The y-axis represents the average number of faces detected; it is
higher than 1 for the original images, because some images have more
than one face.
\ppis \emph{completely foils face detection} for thresholds below 20;
at thresholds higher than about 35, faces are occasionally detected in
some images.

\begin{figure}
    \centering
    \includegraphics[viewport=15 0 400 170, width=3.2in, clip=true]{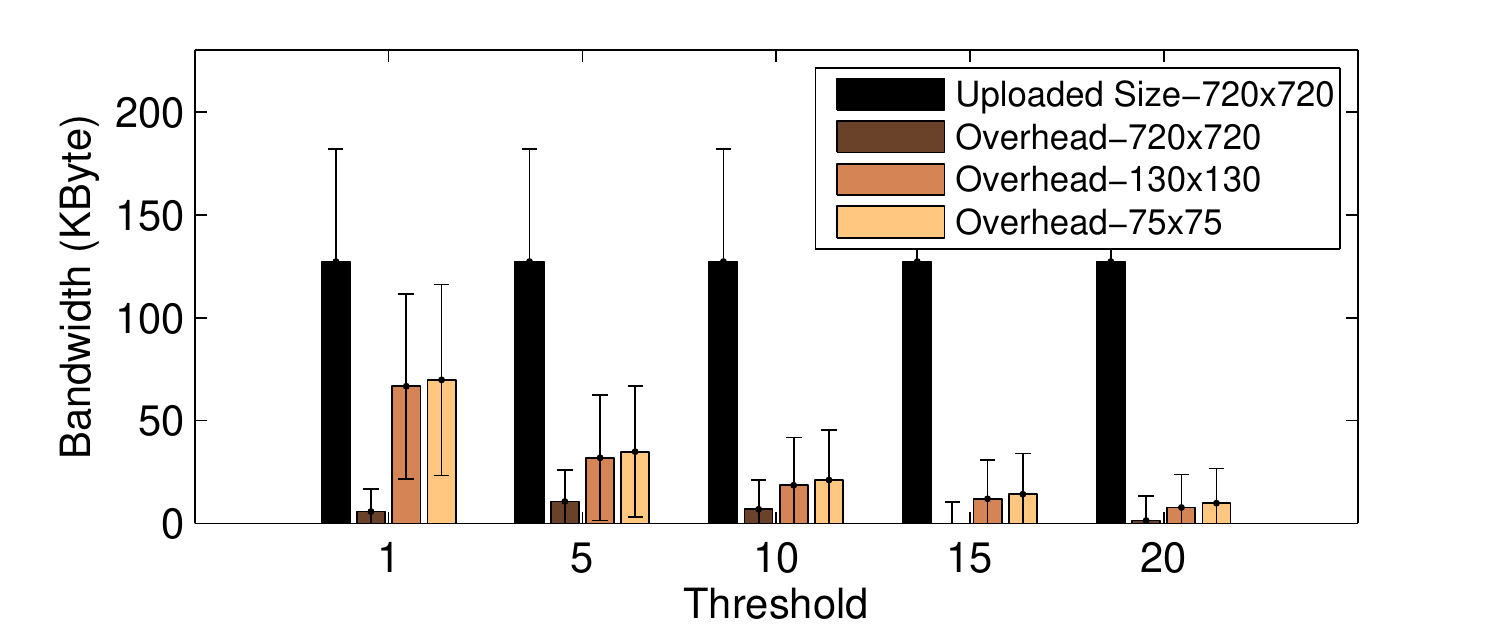}
	\vspace{-5ex}
    \caption{Bandwidth Usage Cost (INRIA)}
    \label{fig:bandwidth_cost}
\end{figure}



\para{SIFT feature extraction.}  
SIFT~\cite{SIFT} (or Scale-invariant Feature Transform) is a general
method to detect features in images.
It is used as a pre-processing step in many image detection and
recognition applications from machine vision.
The output of these algorithms is a set of feature vectors, each of
which describes some statistically interesting aspect of the image.

We evaluate the efficacy of attacking \ppis by performing SIFT feature
extraction on the public part.
For this, we use the implementation~\cite{SIFTimpl} from the designer
of SIFT together with the default parameters for feature extraction
and feature comparison.
Figure~\ref{fig:algo_sift_tradeoff_usc} reports the results of
running feature extraction on the USC-SIPI dataset.\footnote{
  The SIFT algorithm is computationally expensive, and the INRIA data
  set is large, so we do not have the results for the INRIA dataset.
  (Recall that we need to compute for a large number of threshold
  values). We expect the results to be qualitatively similar.
}
This figure shows two lines, one of which measures the total number of
features detected on the public part as a function of threshold.
This shows that as the threshold increases, predictably, the number of
detected features increases to match the number of features detected
in the original figure.
More interesting is the fact that, below the threshold of 10, \emph{no
SIFT features are detected}, and below a threshold of 20,  only about
25\% of the features are detected.

However, this latter number is a little misleading, because we found
that, in general, SIFT detects \emph{different} feature vectors in the
public part and the original image.
If we count the number of features detected in the public part, which
are less than a distance $d$ (in feature space) from the nearest
feature in the original image (indicating that, plausibly, SIFT may
have found, in the public part, of feature in the original image), we
find that this number is far smaller; up to a threshold of 35, a very
small fraction of original features are discovered, and even at the
threshold of 100, only about \camera{4\%} of the original features 
have been discovered.
We use the default parameter for the distance $d$ in the SIFT
implementation; changing the parameter does not change our
conclusions.\footnote{Our results use a distance parameter of 0.6
  from~\cite{SIFTimpl}; we used 0.8, the highest distance parameter
  that seems to be meaningful (~\cite{SIFT}, Figure 11) and the
  results are similar. 
}

%
%
%

\camera{
\para{Face Recognition.}  
Face recognition algorithms take an aligned and normalized face image
as input and match it against a database of faces.
They return the best possible answer, e.g., the closest match or an
ordered list of matches, from the database.
We use the Eigenface~\cite{EigenFaces} algorithm and a well-known face
recognition evaluation system~\cite{CSUFaceRec} with the Color FERET
database.
On EigenFace, we apply two distance metrics, the Euclidean and the
Mahalinobis Cosine~\cite{CSUuserguide03}, for our evaluation.

We examine two settings: \textit{Normal-Public} setting considers the
case in which training is performed on normal training images in the
database and testing is executed on public parts.
The \textit{Public-Public} setting trains the database using public
parts of the training images; this setting is a stronger attack on \ppis 
than \emph{Normal-Public}.
%


Figure~\ref{fig:algo_facerec_feret_fafb} shows a subset of our
results, based on the Mahalinobis Cosine distance metric and using the
FAFB probing set in the FERET database.
To quantify the recognition performance, we follow the methodology
proposed by \cite{Phillips00feret, Phillips98feret}.
In this graph, a data point at $(x,y)$ means that $y$\% of the time,
the correct answer is contained in the top $x$ answers returned by the
EigenFace algorithm.
In the absence of \ppis (represented by the \emph{Normal-Normal}
line), the recognition accuracy is over 80\%.

If we consider the proposed range of operating thresholds (T=1-20),
the recognition rate is below 20\% at rank 1.
Put another way, for these thresholds, more than 80\% of the time, the
face recognition algorithm provides the wrong answer (a false positive).
Moreover, our maximum threshold (T=20) shows about a 45\% rate at rank
50, meaning that less than half the time the correct answer lies in
the top 50 matches returned by the algorithm.
We also examined other settings, e.g., Euclidean distance and other
probing sets, and the results were qualitatively similar. 
These recognition rates are so low that a face recognition attack on
\ppis is unlikely to succeed; even if an attacker were to apply face
recognition on \ppis, and even if the algorithm happens to be correct
20\% of the time, the attacker may not be able to distinguish between
a true positive and a false positive since the public image contains
little visual information.
}

\subsection{What is Lost?}

\ppis achieves privacy but at some cost to
reconstruction accuracy, as well as bandwidth and processing overhead.
%
%



\mypar{Reconstruction Accuracy.}
As discussed in Section~\ref{sec:approach}, the reconstruction of an
image for which a linear transformation has been applied should, in
theory, be perfect.
In practice, however, quantization effects in JPEG compression can
introduce very small errors in reconstruction.
Most images in the USC-SIPI dataset can be reconstructed, when the
transformations are known a priori, with an average PSNR of 49.2dB.
In the signal processing community, this would be considered
practically lossless.
More interesting is the efficacy of our reconstruction of Facebook and
Flickr's transformations.
In Section~\ref{sec:sysarch}, we described an exhaustive parameter
search space methodology to \emph{approximately} reverse engineer
Facebook and Flickr's transformations.
Our methodology is fairly successful, resulting in images with PSNR of
34.4dB for Facebook and 39.8dB for Flickr.
To an untrained eye, images with such PSNR values are generally
blemish-free.
Thus, using \ppis does not significantly degrade 
the accuracy of the reconstructed images.







\mypar{Bandwidth usage cost.}
In \ppis, suppose a recipient downloads, from a PSP, a resized version
of an uploaded image\footnote{\camera{In our experiments, we mimic PSP
   resizing using ImageMagick's convert program~\cite{ImageMagick}}}.
The total bandwidth usage for this download is the size of the resized
public part, together with the complete secret part.
Without \ppis, the recipient only downloads the resized version of the
original image.
In general, the former is larger than the latter and the difference
between the two represents the bandwidth usage cost, an important
consideration for usage-metered mobile data plans.
This cost, as a function of the \ppis threshold, is shown in
Figure~\ref{fig:bandwidth_cost} for the INRIA dataset (the USC dataset
results are similar).
For thresholds in the 10-20 range, this cost is modest: 20KB or less
across different resolutions (these resolutions are the ones Facebook
statically resizes an uploaded image to).
As an aside, the variability in bandwidth usage cost represents an
opportunity: users who are more privacy conscious can choose lower
thresholds at the expense of slightly higher bandwidth usage.
%
Finally, \camera{we} observe that this additional bandwidth usage can be
reduced by trading off storage: a sender can upload multiple encrypted
secret parts, one for each known static transformation that a PSP
performs.
We have not implemented this optimization.

\mypar{Processing Costs.}
On a Galaxy S3 smartphone, for a 720x720 image (the largest resolution
served by Facebook), it takes on average $152$ ms to extract the
public and secret parts, about $55$ ms to encrypt/decrypt the secret
part, and $191$ ms to reconstruct the image.
These costs are modest, and unlikely to impact user 
experience.




\section{Related Work}
\label{sec:related}

We do not know of prior work that has attempted to address photo
privacy for photo-sharing services.
Our work is most closely related to work in the signal processing
community on image and video privacy.
Early efforts at image privacy introduced techniques like
region-of-interest masking, blurring, or
pixellation~\cite{dufaux2010framework}.
In these approaches, typically a face or a person in an image is
represented by a blurred or pixelated version; as ~\cite{dufaux2010framework}
shows, these approaches are not particularly effective against
algorithmic attacks like face recognition.
A subsequent generation of approaches attempted to ensure privacy for
surveillance by scrambling coefficients in a manner qualitatively
similar to \ppis's algorithm~\cite{dufaux2010framework,EbrahimiTalk},
\camera{e.g., some of them randomly flips the sign information.}
However, this line of work has not explored designs under the
constraints imposed by our problem, namely the need for JPEG-compliant
images at PSPs to ensure storage and bandwidth benefits, and the
associated requirement for relatively small secret parts.

This strand is part of a larger body of work on selective encryption
in the image processing community.
%
%
This research, much of it conducted in the 90s and early 2000s, was
motivated by ensuring image secrecy while reducing the computation
cost of encryption~\cite{SE-Survey-Massoudi,SE-Survey-Liu}.
This line of work has explored some of the techniques we use such as
extracting the DC components~\cite{Tang96a} and encrypting the sign of
the coefficient~\cite{Shi98a,Wu00fastencryption}, as well as
techniques we have not, such as randomly permuting the
coefficients~\cite{Tang96a,Qiao97a}.
Relative to this body of work, \ppis is novel in being a selective
encryption scheme tailored towards a novel set of requirements,
motivated by photo sharing services.
In particular, to our knowledge, prior work has not explored selective
encryption schemes which permit image reconstruction when the
unencrypted part of the image has been subjected to transformations
like resizing or cropping. 
%
Finally, a pending patent application by one of the
co-authors~\cite{ortega2002method} of this paper, includes the idea of
separating an image into two parts, but does not propose the \ppis
algorithm, nor does it consider the reconstruction challenges
described in Section~\ref{sec:approach}.

\iftechrep
\camera{
Some recent papers have examined complementary image security and
privacy problems.
Johnson \etal discuss homomorphic encryption based methods for
verifying image signatures when images have been subject to
transformations like cropping, scaling, and JPEG-like
compression~\cite{Johnson2012homomorphic}. 
End-to-end image encryption has been explored for the JPEG 2000 image
format~\cite{Engel2009survey}, and has resulted in a standard for
JPEG 2000 imaging (JPSEC)~\cite{JPSEC}.
}
\fi

Tangentially related is a body of work in the computer systems
community on ensuring other forms of privacy: 
\camera{secure distributed storage systems}~\cite{SPORC,Depot,Depsky}, 
and privacy and anonymity for
mobile systems~\cite{TaintDroid,VirtualTripLine,Anonysense}.
None of these techniques directly apply to our setting.

%
%

%
%



\section{Conclusions}

\ppis is a privacy preserving photo sharing scheme that leverages the
sparsity and quality of images to store most of the information in an
image in a secret part, leaving most of the volume of the image in a
JPEG-compliant public part, which is uploaded to PSPs.
\ppis's public parts have very low PSNRs and are robust to edge
detection, face detection, or sift feature extraction attacks.
These benefits come at minimal costs to reconstruction accuracy,
bandwidth usage and processing overhead.

\camera{ \mypar{Acknowledgements.} We would like to thank our shepherd
  Bryan Ford and the anonymous referees for their insightful
  comments. This research was sponsored in part under the
  U.S. National Science Foundation grant CNS-1048824. Portions of the
  research in this paper use the FERET database of facial images
  collected under the FERET program, sponsored by the DOD Counterdrug
  Technology Development Program Office~\cite{Phillips98feret,
    Phillips00feret}.  }


{
\small
\bibliographystyle{abbrv}
\vspace{1ex}
\bibliography{mybib}
}

\end{document}